\documentclass{aa}  

\usepackage{graphicx}
\usepackage[dvipsnames]{xcolor}
\usepackage{txfonts}
\usepackage[colorlinks,citecolor=blue,bookmarks=false,hypertexnames=true]{hyperref} 
\usepackage{amsmath}
\usepackage{hyperref}

\newcommand{\msun}{\mathrm{M}_\odot}
\newcommand{\mstar}{M_\mathrm{star}}

\graphicspath{{./Figures}}

\begin{document} 
   \title{Galaxies in the simulated cosmic web}
   \titlerunning{Filaments in EAGLE and TNG100}
   \subtitle{I. Filament identification and their properties}

   \author{Yannick M.~Bah\'{e}\thanks{E-mail: yannick.bahe@nottingham.ac.uk}
          \inst{1, 2}
          \and
          Pascale Jablonka
          \inst{1,3}
          }
   
   \institute{
   Laboratory of Astrophysics, \'{E}cole Polytechnique F\'{e}d\'{e}rale de Lausanne (EPFL), Observatoire de Sauverny, 1290 Versoix, Switzerland
   \and
   School of Physics and Astronomy, University of Nottingham, University Park, Nottingham NG7 2RD, UK
   \and 
   GEPI, Observatoire de Paris, Université PSL, CNRS, 5 Place Jules Janssen, 92190 Meudon, France
   }

   \date{Submitted 7 February 2025}

  \abstract
  {As the environment harbouring the majority of galaxies, filaments are thought to play a key role in the co-evolution of galaxies and the cosmic web. In this first part of a series to understand the link between galaxies and filaments through cosmological simulations, we address two major current obstacles on this path: the difficulty of meaningful filament identification, and their poorly constrained properties and internal structure. We use the public EAGLE and TNG100 simulations to build physically motivated filament catalogues with the DisPerSE algorithm, based on the dark matter (DM) field at redshift $z = 0$ and $z = 2$,  explicitly accounting for the multi-scale nature of filaments and with careful validation of results. Filament widths, lengths, and densities vary by factors $\approx\,$5--100 in both simulations, highlighting the heterogeneous nature of filaments as a cosmic environment. All filaments are relatively thin, with overdensity profiles of galaxies, DM, and gas dropping to the cosmic mean within $<\,$3 Mpc from their spines. Contrary to groups and clusters, filament cores are highly substructure dominated, by as much as $\approx\,$80 per cent. Filament gas maps reveal rich temperature and density structures that limit the applicability of simple cylindrically symmetric models. EAGLE and TNG100 agree that $z = 2$ filament spines are traced by overdense cool gas in pressure equilibrium with a $>$10$\times$ hotter envelope. However, significant differences in detail between their predicted gas property maps imply that individual simulations cannot yet describe the baryon structure of filaments with certainty. Finally, we compare our fiducial filament network to one constructed from galaxies. The two differ in many aspects, but the distance of a galaxy to its nearest galaxy-based filament still serves as a statistical proxy for its true environment.}

\maketitle

\section{Introduction}
\label{sec:intro}

On scales of $\gtrsim\,$1 Mpc, the matter distribution in the Universe traces a complex structure that is now commonly referred to as the ``cosmic web'' \citep{Bond1996}: under-dense voids fill the majority of cosmic volume; sheet-like walls surround their edges; one-dimensional filaments lie at the intersection of walls; and finally nodes---massive collapsed haloes that represent the densest cosmic regions---sit at the intersection of filaments. Our view of this large-scale structure emerged from the first systematic galaxy redshift survey (CfA; \citealt{Davis1982}) combined with the theory of anisotropic gravitational collapse \citep{Zeldovich1970}, evidence for the abundance of non-baryonic dark matter (DM; see \citealt{Peebles2017} and references therein), and pioneering numerical simulations of a cold dark matter universe \citep{Davis1985}. More recent surveys have mapped the cosmic web out to at least redshift $z \sim 1$ (e.g.~\citealt{Kitaura2009,Malavasi2017a,Davies2018}), while simulations have confirmed that its structure holds in the concordance $\Lambda$ Cold Dark Matter ($\Lambda$CDM) cosmology (e.g.~\citealt{Springel2005Millennium,Angulo2012}), in the presence of baryons (e.g.~\citealt{Pakmor2023,Schaye2023}), and also when considering alternative dark matter models (e.g.~\citealt{Schneider2012}).

The cosmic web is not just a valuable tool for cosmology (e.g.~\citealt{Bond1996, Amendola2018, Bonnaire2022, Wu2024}),
but is also of central importance for understanding galaxy evolution. The earliest evidence for a link between the two was the overabundance of early-type and red galaxies in regions of higher galaxy density, now commonly referred to as the morphology-density (e.g.~\citealt{Oemler1974,Dressler1980,Postman1984,Bamford2009}) and colour-density relations (e.g.~\citealt{Poggianti1999,Balogh2004,Peng2010}). Closely related trends have since been found for more physical galaxy properties (see e.g.~the reviews by \citealt{BlantonMoustakas2009} and \citealt{Alberts2022}, and references therein). Together, they imply a significant co-evolution of galaxies and their surrounding large-scale structure. This bond is strongest in the densest environments---nodes that correspond to massive haloes---but it persists to the more modest overdensities (e.g.~\citealt{Peng2010,Alberts2022}) that encompass the majority of mass and galaxies in the Universe and are characteristic of filaments (e.g.~\citealt{Hahn2007, Cautun2014}).

While the existence of cosmic filaments is evident from both galaxy surveys and the dark matter distribution predicted by cosmological simulations, detecting them in an objective and quantitative way is considerably more challenging than for haloes. One reason for this is that filaments are generally less dense than haloes, and so stand out less prominently from their surroundings. Another is that filaments are not virialised and therefore lack the (approximate) spherical symmetry of haloes. As a result, there is no analogue to the virial radius that could define a natural boundary of filaments based on analytic theory. Finally, their inherently connected nature makes it difficult to partition the cosmic filament network into its individual elements---again in contrast to haloes, which are typically well separated from each other.

In response to this challenge, a number of sophisticated filament finders have been developed over the last two decades. They are based on a variety of mathematical and / or physical concepts, including stability criteria from the theory of dynamical systems (e.g.~\citealt{Hahn2007}), graph theory (e.g.~\citealt{Barrow1985,Bonnaire2020}), Hessian-based geometry (e.g.~\citealt{Aragon-Calvo2007, Cautun2013}), percolation theory (e.g.~\citealt{Einasto2018}), stochasticity (e.g.~\citep{Tempel2016}), topology (e.g.~\citep{Sousbie2011a}), or phase space (e.g.~\citep{Falck2012}). \citet{Libeskind2018} present an in-depth discussion and comparison of twelve of these codes, all of which use similar terminology for different definitions of filaments.

A sizeable number of works have used these algorithms for quantitative filament studies. Theoretical investigations have analysed filaments in cosmological simulations. They extract the filaments, and potentially other cosmic web environments, either directly from the (dark) matter density field (e.g.~\citealt{Ganeshaiah2019, Martizzi2019}), or from the simulated galaxies in an attempt to match observations (e.g.~\citealt{Dubois2014,Galarraga-Espinosa2020,Hasan2024}). A subset of these works focused on the overall structure and mass content of the filaments, such as their length and width (e.g.~\citealt{Gheller2016,Galarraga-Espinosa2020}), different gas phases (e.g.~\citealt{Martizzi2019,Martizzi2020,Tuominen2021}), their connection to galaxy clusters (e.g.~\citealt{Kuchner2021,Vurm2023, Rost2024}), or their evolution with redshift (e.g.~\citealt{Galarraga-Espinosa2024,Wang2024}). Many others have studied the distribution and properties of galaxies in filaments (e.g.~\citealt{Kraljic2019,Singh2020,Bulichi2024, Hasan2024}).

At the same time, filaments have also been extracted from observed galaxy catalogues with precise distance measurements. Examples include spectroscopic redshift surveys, such as the SDSS (e.g.~\citealt{Malavasi2020,Winkel2021,OKane2024}) or VIPERS \citep{Malavasi2017a}; photometric surveys with high-quality photometric redshifts, such as COSMOS (e.g.~\citealt{Kraljic2018,Laigle2018}); or direct distance measurements in the local Universe (e.g.~\citealt{Castignani2022a,Hoosain2024}). These works have generally focused on the galaxy properties within filaments, in particular their stellar mass, star formation activity, metallicity, and atomic hydrogen content. Most of these works report significant correlations of these properties with the distance between the galaxy and a filament spine.

Despite this large body of literature, we still lack a clear physical understanding of the structure of filaments and how they co-evolve with their galaxies. There are a number of interconnected reasons for this. First, the common approach of identifying filaments from galaxies leads to difficulties in the interpretation of results. For example, massive galaxies are typically surrounded by a large number of `satellites' that create local galaxy density peaks. Filament spines will therefore by construction pass close to them, potentially inducing the observed trend with stellar mass and correlated galaxy properties. Using simulations, \citet{Laigle2018} report broad agreement between dense filament spines identified from galaxies and from DM, but with offsets that can exceed 1 Mpc. In observations, the use of galaxies as filament tracers is currently unavoidable, as the direct detection of diffuse gas in filaments is typically limited to rare individual cases (e.g.~\citealt{Cantalupo2019,Tornotti2024}) or stacks of previously identified filaments (e.g.~\citealt{DeGraaff2019,Tanimura2022}). Disentangling cause and effect in the observed trends is therefore far from trivial.

A second complication is the heterogeneous selection of filaments in the literature. Many works study filaments on the outskirts of galaxy clusters, both in observations (e.g.~\citealt{Malavasi2020,Castignani2022a}) and simulations (e.g.~\citealt{Gouin2022,Vurm2023}). While it is easiest to identify filaments in this environment, their selection is then necessarily biased and it is unclear to what extent results apply to the full population. Similarly problematic is that most studies do not differentiate between different types of filaments and implicitly treat them as one homogeneous environment. This is a major oversimplification (see e.g.~\citealt{Galarraga-Espinosa2020}), akin to neglecting the differences between a dwarf galaxy and a massive elliptical.

Finally, the properties of filaments depend sensitively on the algorithm used to identify them \citep{Libeskind2018,Rost2020}. As an example, the comparison of \citet{Libeskind2018} found a factor $\approx\,$30 scatter in the median DM overdensity of filaments extracted from the same simulation by twelve different codes. The main reason for this striking inconsistency is that the different algorithms were developed for a variety of scientific objectives, and hence all identify subtly different structures as filaments. Most algorithms also contain user-adjustable parameters that have a strong impact on the result (see e.g.~\citealt{Galarraga-Espinosa2024}), but for which there is no commonly-agreed default. When using galaxies as filament tracers, their completeness limit likewise influences the outcome: shallow surveys are limited to relatively massive galaxies (e.g.~$\mstar \gtrsim 10^{10}\,\msun$ for the SDSS) that can only trace the thickest filaments.

In light of these obstacles, progress requires four key ingredients. First, filaments must be identified independent of galaxies to avoid circular conclusions. Second, we must analyse a representative part of the cosmic web to sample the full filament population. Third, individual filaments must be separated to allow grouping them by their properties and thereby account for their heterogeneous nature. Finally, the filament extraction must be characterized by calibrating it to the underlying density field, so that it is clear what type of filaments are identified. To our knowledge, no previous study satisfies all these criteria.

This article is the first in a series aimed at overcoming the aforementioned limitations in order to better understand how the cosmic web and its galaxies influence each other. In the framework of the concordance $\Lambda$CDM cosmology, we want to illuminate the structure of filaments and the role of galaxies within them; the extent to which filaments affect the observed properties of galaxies; how present-day filaments and their galaxies have assembled; and how the star formation history of galaxies is driven by the evolution---or lack thereof---of their large-scale environment, over all of cosmic history. All of these goals require access to the three-dimensional distribution and evolution of dark matter, gas, and galaxies, so that an analysis based on cosmological simulations is the only viable approach.

Modern simulations are well-converged in their predictions of the DM structure on $\gtrsim$Mpc scales (e.g.~\citealt{Schneider2016,Angulo2022}). Over the last decade, they have also made immense progress on baryon properties, especially due to the now common approach of calibrating uncertain simulation parameters to key galaxy properties (for recent reviews, see \citealt{Vogelsberger2020} and \citealt{Crain2023}). A number of independent simulations now predict galaxy populations with broadly realistic masses, sizes, colours, and star formation rates---at least for stellar masses $10^9 \lesssim \mstar \lesssim 10^{11}\,\msun$---and therefore provide a valuable tool to explore the connection between galaxies and large-scale structures. Two prominent examples are EAGLE \citep{Schaye2015} and IllustrisTNG \citep{Nelson2018}. We will analyse these two simulations alongside each other to separate robust predictions from more uncertain ones.

Several authors have previously identified filaments in both of these simulations (e.g.~\citealt{Ganeshaiah2019, Martizzi2019, Galarraga-Espinosa2020, Bulichi2024, Hasan2024}). However, none of the catalogues from these works satisfy our needs: they all either use the positions of galaxies as basis, do not explicitly identify individual filaments and/or their spines, or considered only one of the two simulations. We therefore begin our investigation, in this paper, by constructing purpose-made filament catalogues based on the considerations discussed above, critically assessing the influence of methodology and parameters. While we postpone an in-depth analysis of the galaxy properties in these filaments to the next part of the series, we examine here the predicted variety of filament lengths, radial density profiles, and widths (the latter defined by contrast to the background density), as well as the redshift evolution between $z = 0$ and 2. We furthermore analyse the properties of gas in and around filaments, in anticipation of its critical role in any interactions with galaxies.

The remainder of this paper is structured as follows. In Sect.~\ref{sec:sims} we summarize the most relevant aspects of the EAGLE and IllustrisTNG simulations, before describing in detail our filament identification procedure in Sect.~\ref{sec:filaments}. We then discuss the properties of filaments at $z = 0$ (Sect.~\ref{sec:z0}) and also at $z = 2$ (Sect.~\ref{sec:z2}). Given its key role in the interaction between filaments and galaxies, we then analyse in some detail the properties of filament gas in Sect.~\ref{sec:gas}, before discussing the impact of identifying filaments from galaxies rather than DM in Sect.~\ref{sec:galaxy-filaments}. We summarize our results and present the salient conclusions in Sect.~\ref{sec:summary}. Throughout, we use a $\Lambda$CDM cosmology with the same parameters as in the underlying simulations: $H_0 = 67.77\, \mathrm{km}\,\mathrm{s}^{-1}\,\mathrm{Mpc}^{-1}$, $\Omega_{\Lambda} = 0.703$, $\Omega_\mathrm{m} = 0.307$ for EAGLE, and $H_0 = 67.74\, \mathrm{km}\,\mathrm{s}^{-1}\,\mathrm{Mpc}^{-1}$, $\Omega_{\Lambda} = 0.6911$, $\Omega_\mathrm{m} = 0.3089$ for IllustrisTNG. We denote comoving and proper lengths with the prefix `c' and `p', respectively, and use `$h$-free' units for lengths and masses except when quoting from other works.

\section{The EAGLE and IllustrisTNG simulations}
\label{sec:sims}

Because we aim to study a representative part of the cosmic web, we require a simulation that is large enough to include structures up to the scale of galaxy clusters, but also has high enough resolution for converged galaxy properties down to mass scales well below the Milky Way, $\mstar \sim 10^9\, \msun$. The former implies a volume $V \gtrsim (100\, \mathrm{cMpc})^3$, the latter a (baryon) mass resolution of $m \lesssim 10^6\,\msun$. These conditions are met by two publicly available simulation suites, which we use for our study: EAGLE and IllustrisTNG. They were developed independently of each other and differ in several key modeling aspects, so that a comparison between them allows us to assess the robustness of our results to modelling assumptions and hence enables separating physical from numerical effects. We summarize their aspects most relevant to our work below and refer to \citet{Schaye2015}, \citet{Vogelsberger2014}, \citet{Weinberger2017}, and \citet{Pillepich2018Model} for full details. 

\subsection{The EAGLE simulation}
\label{sec:sims:eagle}

The EAGLE project \citep[see also \citealt{Crain2015}]{Schaye2015} consists of a suite of cosmological $N$-body + Smoothed Particle Hydrodynamics (SPH) simulations, performed with a heavily modified version of the \textsc{Gadget3} code (last described by \citealt{Springel2005}). We mainly use the largest-volume realization of the `Reference' model (Ref-L0100N1504 in \citealt{Schaye2015}; EAGLE[-Ref100] in the following) that evolves a (100 cMpc)$^3$ volume from initial conditions at $z = 127$ to the present day ($z = 0$). Apart from the gravity and hydrodynamics calculation, the simulation includes `sub-grid' prescriptions for relevant astrophysical processes that arise on unresolved scales: gas cooling and heating \citep{Wiersma2009a}, reionization \citep{Schaye2015}, star formation \citep{Schaye_DallaVecchia2008}, metal return from stellar evolution \citep{Wiersma2009b}, energy feedback from star formation in stochastic thermal form \citep{DallaVecchia_Schaye_2012} with an efficiency that scales with local gas density and metallicity \citep{Crain2015,Schaye2015}, as well as supermassive black hole seeding, growth, and the associated energy feedback (`AGN feedback'; \citealt{Rosas-Guevara2015,Schaye2015}). Relevant to our discussion of filament gas properties (Sect.~\ref{sec:gas}), the simulation has a numerical entropy floor to prevent the formation of an inadequately modelled dense-cold gas phase. As a result, temperatures of dense gas ($n_\mathrm{H} \gtrsim 0.1$ cm$^{-1}$) are limited to a temperature floor $T > 10^4$ K that increases with density \citep[see also \citealt{Schaye_DallaVecchia2008}]{Schaye2015}.

As explained by \citet{Schaye2015}, key parameters of these sub-grid models -- in particular the energy efficiency of star formation and AGN feedback -- are poorly constrained and must therefore be calibrated. For EAGLE, this calibration was primarily performed against the observed stellar masses and sizes of galaxies in the local Universe \citep{Crain2015}. This approach also leads to broadly realistic predictions of star formation rates and quenched fractions \citep{Schaye2015}, neutral gas content \citep{Bahe2016, Crain2017}, and galaxy colours \citep{Trayford2015}, amongst many others. The properties of filaments, however, were not considered in this calibration and can therefore be regarded as genuine simulation predictions that allow strong tests of the underlying models.

EAGLE-Ref100 models the evolution of $1504^3$ DM particles (each with mass $m_\mathrm{DM} = 9.69 \cdot 10^{6}\,\msun$) and an initially equal number of gas particles\footnote{A few per cent of these are converted to star or black hole particles, or swallowed by the latter, over the course of the simulation.} (with initial mass $m_\mathrm{gas}^\mathrm{init} = 1.81\cdot 10^6\,\msun$), whose properties are stored at 29 snapshots between $z = 20$ and $0$. In each of these snapshots, structures are identified in a two-step procedure. First, a friends-of-friends (FoF) algorithm with a linking length $b = 0.2$ times the mean inter-particle separation is run on the DM particles to identify spatially disjoint `haloes'. In a second step, the \textsc{Subfind} algorithm \citep[see also \citealt{Springel2001}]{Dolag2009} is used to find gravitationally self-bound `subhaloes' within each FoF halo, now including baryon particles. For this work, we identify subhaloes with a total stellar mass of at least $10^8\,\msun$ as galaxies. At $z = 0$ ($2$) there are 40\,312 (35\,149) galaxies with stellar mass $\mstar > 10^8\,\msun$ in EAGLE-Ref100; 13\,300 (8\,043) have $\mstar > 10^9\,\msun$ and hence reasonably well converged baryonic properties, in particular quenched fractions \citep{Schaye2015}. All data is publicly available \citep{McAlpine2016}.

\subsection{The TNG100 simulation}
\label{sec:sims:tng}

In addition to EAGLE-Ref100, we perform our analysis on the public IllustrisTNG100 simulation (`TNG100' from now on) from the IllustrisTNG project \citep{Marinacci2018, Naiman2018, Nelson2018, Pillepich2018, Springel2018}. This simulation is based on a fundamentally different hydrodynamics solver---the \textsc{arepo} moving mesh code \citep{Springel2010} that also includes ideal magneto-hydrodynamics---and different sub-grid models than EAGLE. In particular, stellar energy feedback is implemented in kinetic from through hydrodynamically decoupled winds with explicitly prescribed speeds and mass loadings, while AGN feedback transitions from a numerically inefficient `quasar mode' to a highly efficient `jet mode' at low black hole accretion rates \citep{Weinberger2017, Weinberger2018}. Similarly to EAGLE, dense gas is limited to temperatures $T \gtrsim 10^4$ K, in this case through an effective equation of state in the sub-grid model for star-forming gas \citep{SpringelHernquist2003}. The key parameters of the TNG model were calibrated against observations of the local Universe that, in contrast to EAGLE, also include galaxy colours and the gas fractions of massive haloes. At the resolution of TNG100, the model reproduces these calibration diagnostics well, in addition to a wide range of other galaxy properties (e.g.~\citealt{Donnari2019, Xu2019, Bustamante2019, Gebek2024}).

The side length of the TNG100 simulation box is $75\,h^{-1} = 110.7$ cMpc, resulting in a 36 per cent larger simulated volume than EAGLE-Ref100. DM particles have a mass of $m_\mathrm{DM} = 7.46 \cdot 10^6\,\msun$, while the average mass of a gas resolution element (i.e.~a gas cell) is $m_\mathrm{gas} = 1.39 \cdot 10^6\,\msun$. Both of these are slightly (23 per cent) lower than for EAGLE-Ref100. The identification of haloes and subhaloes is very similar as for EAGLE, also using the \textsc{Subfind} algorithm. Commensurate with its larger volume, TNG100 contains slightly more galaxies than EAGLE: at $z = 0$ ($2$) there are 49 914 (48 325) with $\mstar > 10^8\,\msun$; 21 052 (12 281) of them have $\mstar > 10^9\,\msun$.

\section{Filament identification}
\label{sec:filaments}

In contrast to haloes and subhaloes, filaments are not detected by \textsc{Subfind} because they are less overdense and not gravitationally self-bound; a dedicated filament finding step is therefore necessary. As we cannot re-use one of the filament catalogues from previous works (see the Introduction), we therefore begin our analysis by building our own, based on the three-dimensional DM density field and the publicly available, well-established topological DisPerSE algorithm \citep{Sousbie2011a, Sousbie2011b}. In the remainder of this section, we first discuss the subtle but important limitations that arise from the multi-scale nature of the cosmic web (Sect.~\ref{sec:filaments:multiscale}), then describe our approach to extract filaments with DisPerSE (Sect.~\ref{sec:filaments:disperse}), and finally discuss our choice for the various parameters inherent in this process (Sect.~\ref{sec:filaments:parameters}). 

\subsection{The problematic multi-scale nature of filaments}
\label{sec:filaments:multiscale}

\begin{figure}
  \includegraphics[width=0.97\columnwidth]{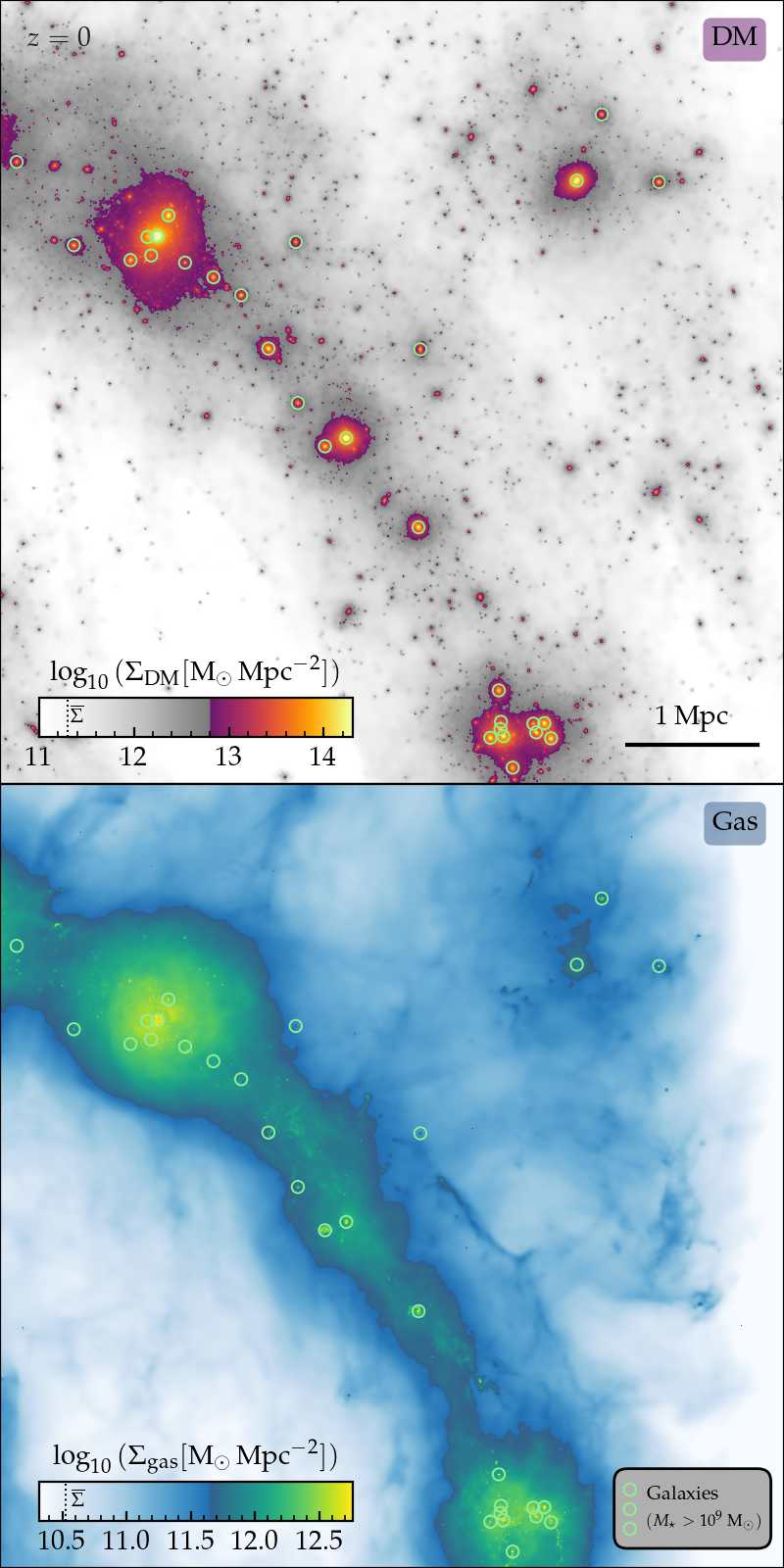} 
  \caption{Close-up view of one filament in the EAGLE-Ref100 simulation. Shown is the dark matter (top) and gas (bottom) surface density in a $6 \times 6 \times 6$ Mpc cube at $z = 0$. In both cases, the colour map changes halfway between the cosmic average ($\overline{\Sigma}$) and the 99.99 percentile, to highlight the low- and high-overdensity regions. Galaxies with $\mstar > 10^9\,\msun$ are depicted as open green circles. Dark matter is concentrated into clumps with a density $> 10 \times$ above the smooth filament background. The gas is more smoothly distributed and shows a continuous filament structure, but still has noticeable peaks up to $\approx\,$5 $\times$ the smooth background, due to the presence of galaxies and feedback-driven haloes. Neither corresponds to the smooth density field implicitly assumed by topological filament finders, which complicates the filament identification.}
  \label{fig:1_filament_close_up}
\end{figure}

The simple picture of the cosmic web as a hierarchical structure of voids, walls, filaments, and nodes only describes the large-scale structure in the Universe to first order. Instead, the real cosmic web is a \emph{multi-scale} structure. Ultra-high-resolution zoom-in simulations have shown that $\Lambda$CDM voids contain plentiful structure in the form of dark matter filaments and haloes, down to at least $\sim$10 pc scales \citep{Wang2020}. As we show in the top panel of Fig.~\ref{fig:1_filament_close_up}, dark matter within filaments is similarly structured into haloes, some of them connected to (smaller) sub-filaments, while much of the main filament has a density closer to the background than to the most prominent peaks. This intricate multi-scale structure arises because filaments, walls, and voids are not virialised and therefore --- unlike within haloes --- the dark matter in them is not stabilised against further gravitational collapse (in fact, we show in Sect.~\ref{sec:z0:substructure} that the substructure fraction is even higher within than outside filaments).

There are two particular consequences of this situation. First, DM filaments are traced by a series of peaks that are relatively isolated locally, and not by a smooth background overdensity like in haloes. This is quite different from what is implicitly assumed by topological filament finders, such as DisPerSE: its approach of identifying filaments as critical lines between peaks and (2-)saddles only works if there is actually a continuous density gradient on the scale of interest. Without corrective steps, the filament network identified by DisPerSE from the DM density field will therefore not correspond to the main underlying structure. For the same reason, the seemingly obvious analogy between filaments and the ridges of a mountain range between two peaks (e.g.~\citealt{Galarraga-Espinosa2024}) captures more their ideal rather than actual nature: mountain ridges \emph{are} locally well-defined, with altitude variations along a ridge typically only a fraction of the peak elevation. A closer topographical analogy for real filaments would be a chain of islands in the ocean.

The second consequence is that there is, strictly speaking, no such thing as a `filament galaxy'. In a $\Lambda$CDM universe, galaxies live inside haloes (e.g.~\citealt{WechslerTinker2018}), which are (local) nodes, and because haloes are the locally dominant environment \emph{even within a filament}, any galaxy that is part of a filament will principally belong to this node rather than to the filament. If we want it to be associated with the latter, we must explicitly discount the nodes represented by their own haloes\footnote{The same argument applies to galaxies within walls and voids, where the local dominance of haloes is even greater.}.

Since both problems stem from the existence of structures below the scale of interest, a natural solution is to filter these `nuisance structures' from the density field before running DisPerSE. This minimum scale of interest must be specified and clearly depends on the scientific objective: kpc-scale filaments that feed individual galaxies, for example, are distracting for our work but of central importance for studies of galaxy connectivity and accretion physics (e.g.~\citealt{Galarraga-Espinosa2023}). Here, we define filaments as relevant if they are thick enough to host well-resolved galaxies ($\mstar > 10^9\,\msun$). This dependence on the science goal means that no filament catalogue is universally applicable, even when one chooses a filament identification method that takes the multi-scale nature of the cosmic web into account (e.g.~NEXUS+; \citealt{Cautun2013}): the choice of relevant scale is then absorbed into the algorithm, but it is implicitly still made.

In practical terms, small-scale structures can be filtered out in one of three ways. First, one may use a sparse tracer---such as galaxies---that effectively hides the multi-scale complexity. As illustrated in Fig.~\ref{fig:1_filament_close_up}, galaxies do indeed line up approximately linearly towards the centre of the DM filament, although by no means exactly. However, in addition to the issue of circular conclusions as discussed in the Introduction, the density field reconstructed from galaxies is necessarily very coarse and ignores much of the useful information provided by the DM: even down to $\mstar = 10^9\,\msun$, there are only 13'300 galaxies in the (100 Mpc)$^3$ volume of EAGLE at $z = 0$, which corresponds to a mean inter-galaxy separation of 4.2 Mpc, and still $\sim$1 Mpc within (overdense) filaments (Fig.~\ref{fig:1_filament_close_up}). Moreover, the scale of identifiable filaments is then not set explicitly by our science goal, but implicitly and indirectly by the availability of galaxies.

Another option is to use a highly-sampled tracer that is less susceptible to small-scale structure formation, i.e.~the gas. As illustrated in the bottom panel of Fig.~\ref{fig:1_filament_close_up}, the gas distribution is indeed much smoother than the DM, with the densest part forming a continuous structure as opposed to the very isolated density peaks for DM. Nevertheless, local density peaks persist at the location of galaxies. More problematically, gas in the vicinity of haloes is subject to strong feedback from star formation and AGN. This not only creates structures that have no counterpart in DM and are hence arguably not part of the cosmic web, but different plausible feedback implementations in simulations lead to strongly differing outcomes in terms of how gas is ejected from haloes (see e.g.~\citealt{VanDaalen2020}). Identifying filaments from the gas would therefore complicate the interpretation of results substantially. 

Given these unsatisfying alternatives, we use a third option: our filament finding is based on the position of simulated DM particles, with the problem of small-scale structure mitigated by smoothing the DM density field before identifying cosmic web features. As we detail below, this approach requires calibration and is therefore, in the absence of a well-defined benchmark,
necessarily somewhat subjective. However, combined with careful examination of the result it allows us to identify cosmic filaments in a quantitative and physically meaningful way.

\subsection{DM filament identification with DisPerSE}
\label{sec:filaments:disperse}

Based on the considerations discussed above, we identify filaments from the simulated DM density field in a sequence of six steps, each involving one or more parameter choices. For clarity, we first describe our fiducial procedure, giving full details but only minimal explanations. We then revisit each step individually in Sect.~\ref{sec:filaments:parameters}, where we explain and justify our choices.

Starting from the coordinates of individual DM particles, we construct the DM density field on a regular three-dimensional grid, with 1\,200 cells per dimension for EAGLE, corresponding to a cell size of 83.3 kpc. For TNG100, we use the same cell size, i.e.~a grid with 1\,329 cells per dimension. We use nearest-grid-point interpolation, in other words assigning the entire mass of a particle to the cell at its position. The resulting density field is then smoothed to suppress small-scale features in the (real) density field, as well as reduce the sensitivity to sampling noise\footnote{We have found that DisPerSE crashes consistently without this step, likely because a large number of cells in the unsmoothed field have a value of zero.}. Following \citet{Cautun2013}, we smooth the \emph{logarithm} of the density field ($\log_{10}\,\rho / \overline{\rho}$, where $\overline{\rho}$ denotes the cosmic mean density). Smoothing is done with a spherical Gaussian kernel with a standard deviation of 0.5 Mpc (see Sect.~\ref{ssec:smoothing}). Once smoothed, we downsample the density cube by a factor of $3^3$, i.e. to $N = 400$ cells of side length $\Delta x = 0.25$ Mpc per dimension. This downsampling is necessary for DisPerSE to run in a reasonable length of time ($\sim$5 hours) and within the constraints of available memory.

With the final DM density cube constructed, we extract the filamentary `skeleton' of the cosmic web with DisPerSE, specifically its \texttt{mse} task. We run \texttt{mse} in its alternative mode with the input density provided as a FITS cube, with fully periodic boundary conditions as appropriate for our simulation input. In this mode, the `persistence threshold' $T$ must be specified as an absolute value, rather than in terms of the statistical significance (`multiples of $\sigma$') typically used with discrete sample inputs such as galaxy positions. As detailed in Sect.~\ref{ssec:persistence}, we choose a threshold $T = 4 \cdot 10^7\,\msun\,/V_\mathrm{cell}$, where $V_\mathrm{cell} = 1.56 \cdot 10^{-2}\, \mathrm{Mpc}^{-3}$ is the volume of one DM field cell, to capture the majority of visually clear and relevant filaments while minimising the number of filament identifications that do not correspond to clear features in the DM field.

DisPerSE traces filaments by their spines, which are made up of short ($\sim$300 kpc) straight-line `segments' between a sequence of `sampling points'. As described by \citet{Sousbie2011b}, the exact location of these sampling points is noisy and suffers from discretisation, which leads to artificial small-scale structure in the identified filaments. To mitigate this, we smooth the filaments with the \texttt{skelconv} routine of DisPerSE, iteratively replacing each sampling point with the average of its two neighbours $N_\mathrm{smooth} = 10$ times (see Sect.~\ref{ssec:filament-smoothing}). The result is a smooth sequence of sampling points, with 80 per cent of adjacent segments differing in their orientation by less than 5 degrees.

To cleanly separate filaments from the (major) nodes of the cosmic web, i.e.~galaxy groups and clusters, we identify any filament sampling points that lie within $3\, r_\mathrm{200c}$ of a FoF halo\footnote{We define $r_\mathrm{200c}$ as the radius of a sphere centred on the potential minimum of the halo within which the average density is 200 times the critical density of the Universe. The mass within this sphere is denoted as $M_\mathrm{200c}$.} with mass $M_\mathrm{200c} > 10^{13}\,\msun$ (see Sect.~\ref{ssec:masking}). Segments connected to any such points are masked out and not considered as part of a filament. In other words, no identified filament spine has any part that lies within $3\,r_\mathrm{200c}$ of a group or cluster.

As a final step, we test whether any two filaments that start or end at the same `critical point' (node or 2-saddle) are sufficiently closely aligned that they should be considered as two parts of the same filament, rather than two separate structures. Since each critical point may be connected to an arbitrary number of filaments, we first identify for each critical point the two closest-aligned filaments; no others are considered for joining. We then join these two filaments if the angle between them is less than 60 degrees (see Sect.~\ref{ssec:joining}), i.e.~if their direction vectors are (in 3D space) closer to being parallel than perpendicular. This criterion is rather permissive and only excludes 11 per cent of potential joins.

\begin{figure*}
  \includegraphics[width=\textwidth]{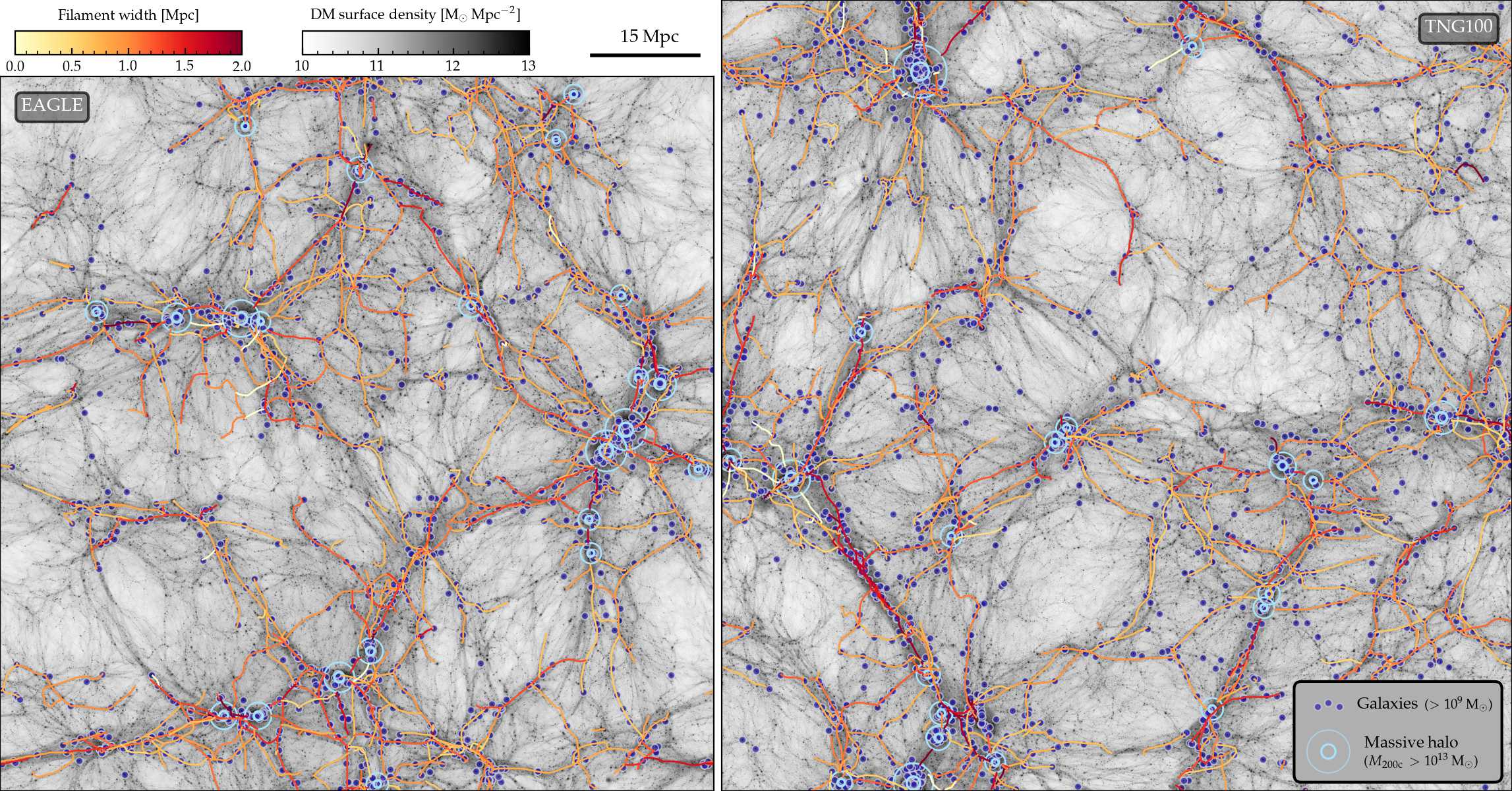}  
  \caption{Filament network at $z = 0$ of EAGLE (left) and TNG100 (right) compared to the underlying DM density field. Orange lines represent filaments in a 15 Mpc thick slice. Background images show the DM density projected in the same slice. Galaxies with $M_\mathrm{star} > 10^9\,\mathrm{M}_\odot$ are shown as indigo circles. The filaments identified by DisPerSE trace well the main filamentary features of the DM density field that harbour resolved galaxies.}
  \label{fig:2_overview}
\end{figure*}

The end result of this procedure is a catalogue of 2\,400 and 3\,189 filaments in EAGLE and TNG100, respectively\footnote{Excluding filaments with a length of zero, which are likely an artefact of DisPerSE.}. A visual impression of them is shown in Fig.~\ref{fig:2_overview}, where we plot the filaments in shades of orange and red over the underlying DM density (greyscale). Both DM and filaments are only shown in a 15 Mpc thick slab to reduce projection effects.

The filaments identified by our approach correspond closely to the major structures of the DM density field. Unsurprisingly, they are particularly prominent, and generally thickest (red shades, see below) between the most massive haloes (light blue circles), but they extend throughout the simulation volumes. There is a plethora of additional DM filaments that we do not identify, but as discussed above (Sect.~\ref{sec:filaments:multiscale}) this is by choice---hardly any of them contain galaxies with $\mstar > 10^9\,\msun$ (indigo circles). Visually, the global structure of the filament networks of EAGLE and TNG100 is very similar. This is not surprising given that the $N$-body part of the \textsc{gadget} and \textsc{arepo} codes is very similar and that, more generally, the predictions for cold dark matter structures are well-converged between simulations (see section 8.5 of the review by \citealt{Angulo2022} and references therein).

\begin{figure}
  \resizebox{\hsize}{!}{\includegraphics{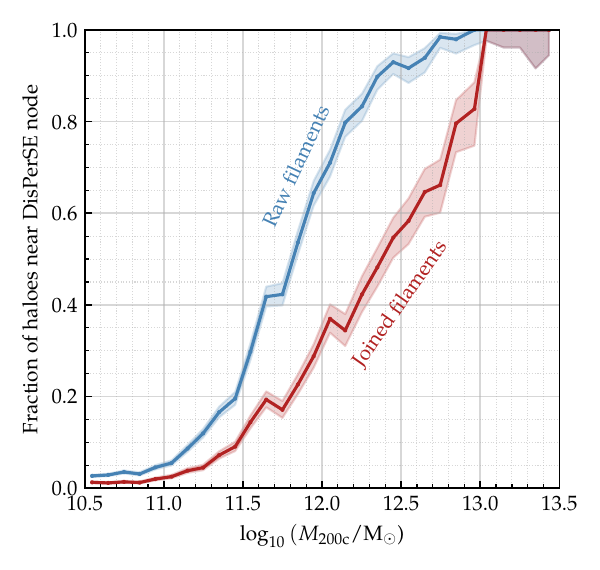}}
  \caption{Fraction of haloes that mark the endpoint of filaments. Shown is the fraction of FoF haloes, in 0.1 dex bins of $M_\mathrm{200c}$, that are within 500 kpc of a cosmic web node as identified by DisPerSE (blue: all nodes, red: only those at the end of a joined filament; see text). Shaded bands represent the binomial 1$\sigma$ uncertainty on the fraction. There is no single mass threshold above which haloes become filament endpoints, but for the joined filaments the transition occurs broadly at the scale of massive galaxies ($M_\mathrm{200c} \approx 2\cdot10^{12}\,\msun$).}
  \label{fig:3_halo_node_connection}
\end{figure}

As a quantitative measure of our cosmic web identification, we show in Fig.~\ref{fig:3_halo_node_connection} the overlap between the nodes as identified by DisPerSE and the DM haloes from the FoF algorithm. While only a few per cent of haloes with $M_\mathrm{200c} < 10^{11}\,\msun$ lie within 500 kpc of a DisPerSE node, the fraction increases steadily towards higher mass and reaches unity at $10^{13}\,\msun$. This relatively broad transition range, over a factor 100 in halo mass, is due to our log-smoothing procedure that effectively prioritises the large-scale structure around a halo, rather than its local mass concentration, in determining its significance---in other words, it confirms that our approach actually identifies filamentary structures rather than just arbitrary connections between massive haloes. 

As should be expected, the end-points of the joined filaments typically correspond to more massive haloes than the individual nodes (red vs.~blue lines in Fig.~\ref{fig:3_halo_node_connection}). They only cover the majority of haloes above a mass of $\approx\,$2$\,\cdot\,10^{12}\,\msun$, but the fraction then rises steeply and also reaches unity at $10^{13}\,\msun$. The latter is by construction, as our masking approach prevents joining filaments at a group/cluster halo.

\subsection{Parameter choices for our filament identification}
\label{sec:filaments:parameters}

Having described our fiducial filament extraction above, we now discuss our choice of approach and parameters in the individual steps. Readers who are mainly interested in the properties of the identified filaments may skip this part and continue to Sect.~\ref{sec:z0}.

\subsubsection{DM density field construction}
\label{ssec:dmfield}
Although DisPerSE can in principle be run directly from the coordinates of a discrete set of sampling points---such as DM particles---we work with the DM density field sampled on a regular grid, for two reasons. First, the very large number of DM particles in our simulations (ca.~3.4 and 6.0 billion for EAGLE and TNG100, respectively) would make the analysis prohibitively expensive. More importantly, this approach allows us greater flexibility in how to smooth the density field before running DisPerSE.

The grid size of $400^3$ cells for EAGLE ($443^3$ for TNG100 due to its $\sim$10 per cent large simulation box length) is chosen as a compromise between resolution and computational complexity. With a cell size of $\Delta x = 250$ kpc, the Mpc-scale structures of the cosmic web that are of interest here are well resolved. Using the EAGLE-Ref25 simulation, whose volume is 64 times smaller than Ref100, we have verified that a four times smaller cell size (i.e.~$\Delta x = 62.5$ kpc) leads to negligible changes in the identified filament network. Nevertheless, it is important to keep in mind that the exact position of our identified filament spines is uncertain below the 250 kpc scale of a single cell.

To construct the density field from the discrete particles, we choose the simplest approach of assigning the mass of each particle entirely to the cell at its coordinates. This implicitly treats the particles as point masses, while they are in reality Monte Carlo tracers of the underlying continuous DM density field. The reason why this approximation is justified here is the relatively high resolution of both EAGLE and TNG100: the mean separation between DM particles is 66 kpc (EAGLE) and 61 kpc (TNG100), several times smaller than our adopted cell size. We have verified explicitly that an alternative interpolation scheme (cloud-in-cells) yields near-identical results.

In practice, we first sample the DM density field on a 3$\times$ finer grid, with $N$ = 1\,200 (1\,329) for EAGLE (TNG100), smooth this high-resolution DM density field, and then bin the result to the final $N=400$ grid. We do this in order to to better resolve the densest parts of the cosmic web during the smoothing, but have not explored the impact of this choice in detail\footnote{Due to the non-mass-conserving nature of our logarithmic smoothing, sampling the DM density directly on an $N = 400$ grid will give a quantitatively different result, although this could plausibly be accounted for by a re-calibration of the DisPerSE persistence threshold.}.

\subsubsection{DM density smoothing}
\label{ssec:smoothing}
As discussed above, the DM density field must be smoothed before running DisPerSE, for both physical and numerical reasons. The simplest approach would be to use a Gaussian kernel with some specified width. However, we have found that this leads to strong spherically symmetric `bleeding' of the highest density peaks into their surroundings for smoothing lengths $S \gtrsim 0.1$ Mpc, with the result that the smoothed field emphasizes rather than suppresses these small-scale haloes.

To circumvent this problem, we follow the solution proposed by \citet{Cautun2013} for their NEXUS+ algorithm and smooth the density field in logarithmic space: we first compute $\rho^\prime = \log_{10}\,(\rho / \overline{\rho}$), smooth that, and afterwards transform back to linear density ($\rho / \overline{\rho} = 10^{\rho^\prime}$). In effect, this approach down-weights the contribution from the highest density cells and therefore results in a smoothed field that is more representative of the lower-density regions, such as filaments (see Fig.~\ref{fig:A1_smoothing_type_comparion} in Appendix \ref{app:logsmoothing} for a visual comparison). For example, at fixed smoothing length a single cell with an overdensity of 100 (i.e.~$\rho^\prime = 2$) will contribute 50 times less to a neighbouring cell when the smoothing is done in logarithmic rather than linear space.

A technical issue with this approach is that many cells are empty, especially in void regions, and hence have no valid value of $\rho^\prime$. One solution would be to impose a very low but non-zero density floor. However, these cells would then have a strongly negative $\rho^\prime$ that would dominate the smoothed field, which is clearly undesirable. At lower grid resolution ($N=400$), most cells do contain particles, so that we can measure a meaningful lowest overdensity, $\rho / \overline{\rho} \approx 1/30$. Assuming that this is also approximately the lowest true density on the scale of the (3$\times$ smaller) cells in the high-resolution cube, we use a slightly lower value than this as our floor, adding $3 \cdot 10^5\,\msun$ to the mass within each cell ($\approx 1/60^\mathrm{th}$ of the mean cell mass) and then subtract the same mass from each cell after smoothing\footnote{This does not lead to negative $\rho$ in the final result because smoothing always increases the density of the lowest-density cells.}. We have verified that using a ten times higher or lower floor leads to a normalised smoothed density field that is nearly indistinguishable from our default choice---the absolute density values differ by a factor of $\approx\,$3 at the low-density end, but this would simply lead to a different choice of DisPerSE persistence threshold.

\begin{figure*}
  \includegraphics[width=\textwidth]{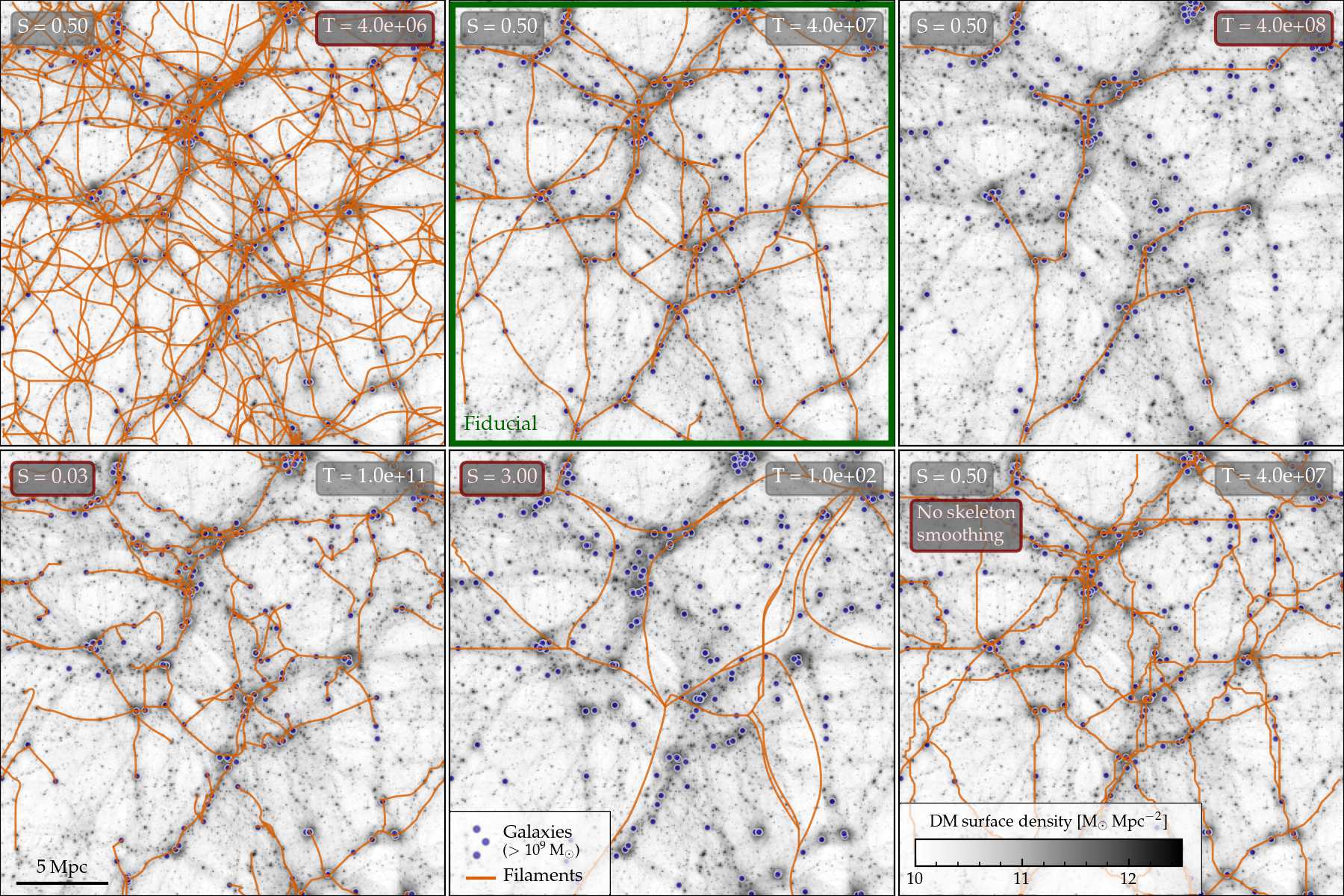} 
  \caption{Sensitivity of the filament network to parameter choices. Each panel shows a full projection through the EAGLE-Ref25 simulation, with filaments in orange, galaxies ($\mstar > 10^9 \msun$) in indigo, and the projected DM density as greyscale image in the background. Deviations from our fiducial parameters are highlighted in red. In the top row, the middle panel (green frame) uses our fiducial values, while the left- and right-hand panels adopt a $10\times$ lower and higher DisPerSE persistence threshold $T$, respectively. In the bottom row, the left-hand and middle panels instead show the effect of smoothing the DM field with a smaller or larger kernel $S$ than our fiducial choice. Finally, the bottom-right panel shows the filaments without smoothing the spines. Of the six networks shown here, only our fiducial choice traces the DM filaments adequately.}
  \label{fig:4_disperse_variations}
\end{figure*}

There is no natural length scale over which to smooth the DM density field. We have therefore calibrated this parameter, such that the resulting filament network traces the key features of the DM field for our science goals as best as possible, again using the smaller EAGLE-Ref25 simulation. We have tested twelve values between $S = 0.03$ and 3 Mpc. For each of them, we have run DisPerSE with many different persistence thresholds to find the optimal value (see below), and then compared the best results across smoothing lengths to find our fiducial value of $S = 0.5$ Mpc. As illustrated in Fig.~\ref{fig:4_disperse_variations}, a smaller value ($S = 0.03$, bottom-left corner) leads to filaments with many small-scale `wiggles' due to local density variations; if it is larger ($S = 3.0$, bottom-middle panel) the filaments no longer follow the underlying DM density.

\subsubsection{DisPerSE persistence threshold}
\label{ssec:persistence}
The relatively high resolution of EAGLE and TNG100 means that even very fine filamentary structures can be statistically significant, down to scales well below that of galaxies. To avoid being swamped by these mini-filaments (even after density smoothing), and to limit the fragmentation of major filaments into many smaller ones, we must therefore calibrate the `persistence threshold' that DisPerSE uses to assess the significance of a filament end-point, based on the level of detail of the resulting filamentary network. Specifically, we want to find the sweet spot between a threshold that is too low and would hence lead to too many insignificant filaments, and one that is too high and results in missing filaments that harbour resolved galaxies ($\mstar > 10^9\,\msun$). We do this in a semi-quantitative way, running DisPerSE with a wide range of thresholds\footnote{To accomplish this large suite of DisPerSE runs in an efficient manner, we use the smaller EAGLE-Ref25 simulation and re-use the previously computed Morse-Smale Complex when changing the persistence threshold.} (from $10^6$ to $10^9$ $\msun$ per cell for our fiducial smoothing length of $S = 0.5$ Mpc) and picking the one for which $\gtrsim 90$ per cent of galaxies lie within 1 Mpc of an identified filament spine, a decrease in the threshold by 25 per cent leads to a $\lesssim$ 1 per cent increase in the covered galaxy fraction, and the filament network covers the majority of visually clear DM filaments with resolved galaxies.

As illustrated in the top row of Fig.~\ref{fig:4_disperse_variations}, the identified filament network is quite sensitive to the choice of persistence threshold: lowering it by a factor of 10 from our fiducial value of $4 \cdot 10^7 \,\msun$ per cell leads to an extremely dense network with filaments everywhere, while a $10 \times$ increase results in a very sparse structure that misses many visually clear filaments that host resolved galaxies. Even though our choice is to some extent subjective, it is therefore physically meaningful and --- in the absence of an objective justification for a particular threshold --- arguably the optimal value for the purpose of classifying filaments as galaxy environments. 

\subsubsection{Filament smoothing}
\label{ssec:filament-smoothing}
Because we run DisPerSE on a relatively coarse regular grid (cell size $\Delta x = 250$ kpc), the filaments as identified by \texttt{mse} are artificially jagged and must be smoothed to correspond better to the actual DM structure (bottom-right panel of Fig.~\ref{fig:4_disperse_variations}). The \texttt{skelconv} task of DisPerSE that we use for this purpose has one free parameter, namely the number $N_\mathrm{smooth}$ of smoothing iterations in which filament sampling points are replaced by the average of their neighbours. As for the DM smoothing length and persistence threshold, we calibrate this number, here by comparing nine filament networks of EAGLE-Ref25 with $N_\mathrm{smooth}$ between 0 and 50. $N_\mathrm{smooth} \gtrsim 10$ is necessary to suppress most of the small-scale wiggles that arise from the discretised input, while there is no appreciable change when going to larger values. We therefore use $N_\mathrm{smooth} = 10$.

\subsubsection{Masking of groups and clusters}
\label{ssec:masking}
Our halo mass threshold for masking ($M_\mathrm{200c} = 10^{13}\,\msun$) is motivated by the observed quenched galaxy fraction as a function of stellar and halo mass (e.g.~\citealt{Wetzel2012}) --- low-mass galaxies ($\mstar \sim 10^{10}\,\msun$) have a quenched fraction excess of $>$25 per cent only in haloes above this threshold. In addition, haloes more massive than $10^{13}\,\msun$ are all identified as a node by DisPerSE (see the blue line in Fig.~\ref{fig:3_halo_node_connection}), whereas the same is not true for lower-mass haloes (e.g.~only two thirds of haloes with $M_\mathrm{200c} = 10^{12}\,\msun$ are identified as a node). A threshold lower than $10^{13}\,\msun$ would therefore have broken up filaments, whereas a higher one would have blurred the distinction between the filament and group/cluster environments.

We choose to mask not only the central part of the group/cluster halo, but also its outskirts within $3\,r_\mathrm{200c}$. This is because a large fraction of galaxies (and hence also DM) within this region has previously passed through the halo centre, disrupting any filamentary structure (see e.g.~\citealt{Bahe2013}). Moreover, the gaseous halo around groups and clusters extends to approximately this radius \citep{Bahe2013,Zinger2018}, so that the environment is dominated by the nearby group/cluster.

\subsubsection{Filament joining}
\label{ssec:joining}
\begin{figure}
  \includegraphics[width=\columnwidth]{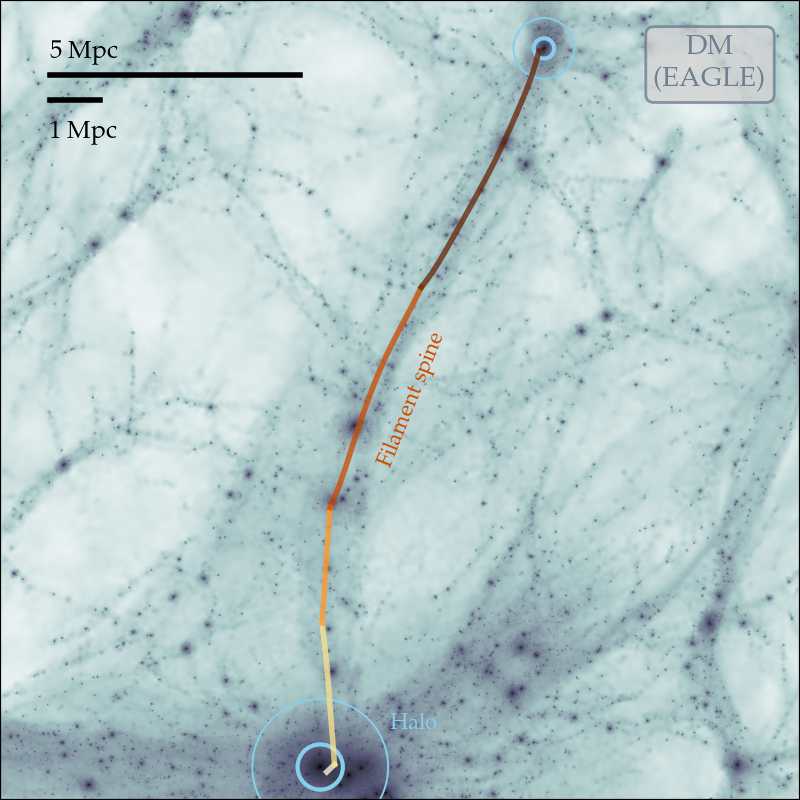} 
  \caption{Comparison of raw and joined filaments. Different coloured lines correspond to five individual (`raw') DisPerSE filaments that we join to one filament for our analysis. The background image shows the projected DM density in a 16 $\times$ 16 $\times$ 4 Mpc slab, oriented along the best-fit plane of the joined filament. Blue circles mark 1 and 3 $r_\mathrm{200c}$ of the two most massive halo in the field, with masses just under $10^{13}$ and $10^{12}\,\msun$, respectively. Linking these two haloes, the joined filament is more physically meaningful than the raw ones.}
  \label{fig:5_joined_filaments}
\end{figure}

As a consequence of our relatively low persistence threshold, large filaments may be artificially split up into multiple smaller ones if there are significant enough local DM concentrations along them (such as moderately massive haloes). An example is shown in Fig.~\ref{fig:5_joined_filaments}, where five `raw' filaments (as identified by DisPerSE) line up smoothly with each other and should therefore be considered as one continuous filament that spans between the two haloes marked with blue circles.

To find the maximum angle between two filaments that should be joined, we consider the distribution of angles $\alpha_\mathrm{seg}$ between segments within each raw filament; we use an unsmoothed version here ($N_\mathrm{smooth} = 0$; see above) because the smoothing procedure does not change the angle between two adjacent filaments systematically while reducing (by design) that between segments within a filament. We find that there is a relatively broad distribution of angles, up to $\alpha_\mathrm{seg} \approx 60$ degrees, while larger angles are much more common between than within filaments. We therefore use 60 degrees as the limit for joining filaments.

We finally note that, while the vast majority of joined filaments have the normal linear shape as depicted in Fig.~\ref{fig:5_joined_filaments}, there are a few cases where filaments curve back onto themselves and end at a point that a point that they have previously traversed (in both EAGLE and TNG100). Because these cases are so rare, and their spines remain locally well-defined, we do not exclude them from our analysis.

\section{The properties of filaments at $z = 0$}
\label{sec:z0}

With the filaments identified as described above, we now characterise their basic properties. The filament length is straightforward to compute as the total length of all its segments, but because DisPerSE only identifies the one-dimensional filament spines, rather than their full volumes, an additional step is necessary to determine their width and density. We derive both quantities from the DM particle distribution around the filaments (Sect.~\ref{sec:z0:widthdef}) and analyse the distribution and correlation of all three properties (Sect.~\ref{sec:z0:properties}). We then compare the overdensity profiles of DM, gas, and galaxies around filaments (Sect.~\ref{sec:z0:multiprofiles}) and quantify the influence of substructures within them (Sect.~\ref{sec:z0:substructure}). 

\subsection{Definition of filament width}
\label{sec:z0:widthdef}

\begin{figure}
  \includegraphics[width=\columnwidth]{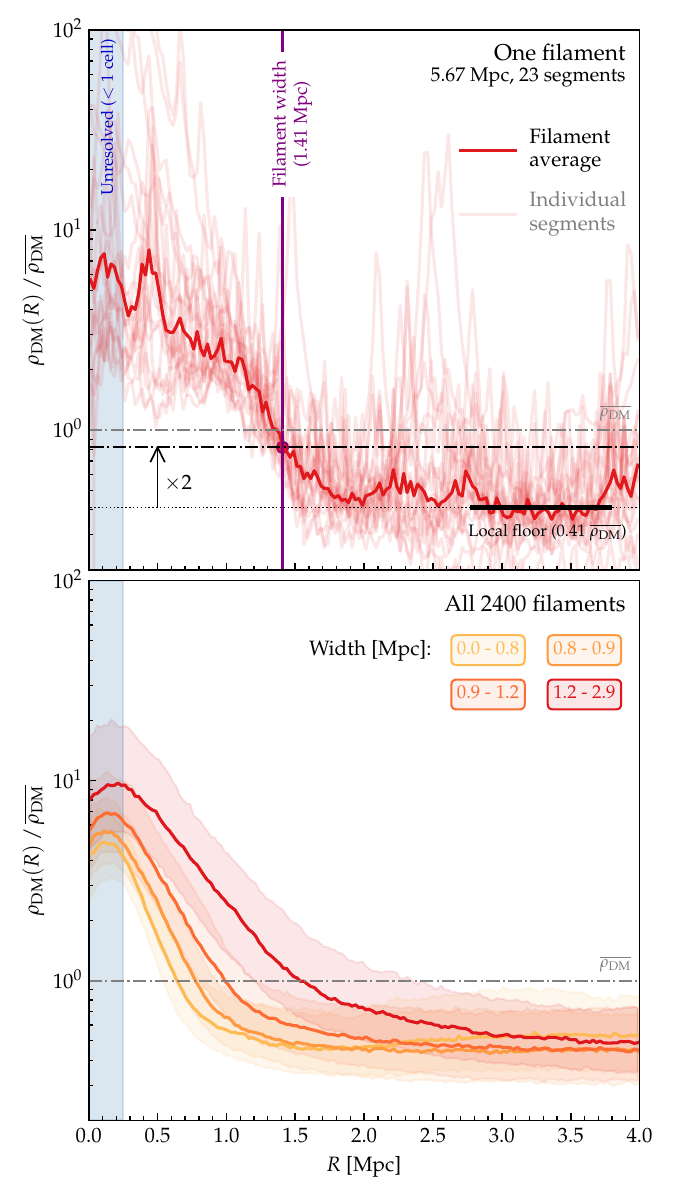}
  \caption{DM density profiles around filaments in EAGLE. \emph{Top:} density profiles for the individual segments of one filament (thin red lines) and their stacked median, the representative profile of the filament (thick red line). The dotted black line represents the local background density, with the solid portion indicating the window where it is measured. The filament width (vertical purple line) is defined as the point where the profile drops below twice this value (black dash-dotted line), which is here close to the cosmic mean (grey dash-dotted line). \emph{Bottom:} Median-stacked profiles of all filaments in bins of their width (different colours); shaded bands enclose the central 50\% of filaments in each bin. All stacks decline steadily to a common floor at around half the cosmic mean, with a clear correlation between the width and central density.}
  \label{fig:6_filament_density_profiles}
\end{figure}

As illustrated in Figs.~\ref{fig:1_filament_close_up} and \ref{fig:5_joined_filaments}, filaments have no sharp edge that unambiguously defines their extent. To determine the filament width, we therefore consider the cylindrical DM density profile around the filament and aim to find the distance from the spine at which this density has dropped sufficiently to identify it as the `boundary' of a filament. Our approach is illustrated in the top panel of Fig.~\ref{fig:6_filament_density_profiles}. For each filament, we first consider all its segments individually and measure the DM density within 160 concentric rings of width 25 kpc, out to 4 Mpc\footnote{Recall that each segment of the spine is defined as a straight line between two sampling points, so that we can measure an unambiguous distance from the spine for each DM particle that lies in the cylinder around each segment. We assign particles wholly to the bin that contains their coordinates.}. The resulting density profiles, in units of the mean DM density $\overline{\rho_\mathrm{DM}}$, are plotted as faint red lines; they show a generally decreasing trend, albeit with significant scatter. We then compute the median of these segment profiles as the characteristic density profile of the filament (thick red line in Fig.~\ref{fig:6_filament_density_profiles}), weighted by the length of each segment. Compared to a mass-weighted mean, this has the advantage of being less susceptible to the presence of relatively massive haloes in a small number of segments that could otherwise dominate the stacked profile (see Fig.~\ref{fig:9_filament_density_profiles} below). In other words, the profile that we construct is representative of the filament along its full length, rather than only its densest parts. 

From this filament DM profile, we then determine the local DM density background $\rho_\mathrm{floor}$ as the minimum of the average density within a sliding window of width 1 Mpc. In Fig.~\ref{fig:6_filament_density_profiles}, we illustrate the resulting background value and the window over which it is measured as a thick horizontal black line, in this case extending from 2.8 to 3.8 Mpc around the filament and with a density of $0.4 \, \overline{\rho_\mathrm{DM}}$. We then find the (smallest) radius at which the profile drops within a factor 2 of $\rho_\mathrm{floor}$, interpolating between adjacent bins, and take this radius as the `width' of the filament\footnote{We refer to the boundary radius as the filament width for simplicity, even though it is more accurately the half-width of the full filament.}. The factor 2 is to some extent arbitrary, but we have found that less than 5 per cent of filament profiles rise again to more than $2\,\rho_\mathrm{floor}$ after having dropped below $\rho_\mathrm{floor}$ (not shown).

In the example shown in Fig.~\ref{fig:6_filament_density_profiles}, the filament width is close to the radius at which the profile drops below $\overline{\rho_\mathrm{DM}}$, which might seem a more natural boundary. We have decided against this because it would introduce an implicit dependency of the filament width on the background: in a large-scale overdensity, the density profile of even a very thin filament may only reach the cosmic mean several Mpc from its spine. By explicitly measuring the local background around each filament, our method is not subject to this bias; we have verified that the filament width is indeed not correlated with the background density (not shown).

In the bottom panel of Fig.~\ref{fig:6_filament_density_profiles}, we show the filament density profiles stacked within the four quartiles of filament width, with yellow representing the thinnest and dark red the thickest filaments; shaded regions indicate the central 50 per cent of filaments in each bin. The stacked profiles show a well-defined decline from the centre, with typical overdensities of $\approx\,$4--10, to a floor of $\approx\,$0.5 $\overline{\rho_\mathrm{DM}}$ at large radii. The slight drop within $\approx,$0.25 Mpc may be physical, but it could plausibly also result from our finite DM cell size leading to a slightly inaccurate placing of the filament spines. The difference between the four bins persists to the centre (i.e.~the region immediately around the filament spine), where the thickest filaments typically have the highest density and vice versa. For clarity, we only show profiles for EAGLE here, but we have verified that TNG100 gives nearly indistinguishable results.

Our overdensity profiles can be compared to those of \citet[top panel of their fig.~40]{Cautun2014} and \citet[top-left panel of their fig.~4]{Zhu2021}, both of which use a tidal tensor approach to find filaments. Despite this fundamental difference from our work, we find a similar dependence on filament width, and broadly agree on the central overdensities of filaments with width $\lesssim\,$3 Mpc. We have no analogues to the thickest profiles of \citet{Cautun2014}, likely because they define filament width locally (on a scale of $4\,h^{-1}$ Mpc), rather than using a constant value for an entire filament as we do. At larger radii, our profiles decline both more rapidly and to a lower asymptote than in either \citet{Cautun2014} or \citet{Zhu2021}. Given the close agreement between EAGLE and TNG100, we suspect that this is not related to differences in the underlying simulations, but instead reflects our different definitions of filaments and how to stack their profiles along the spine.

\subsection{Filament widths, lengths, and densities}
\label{sec:z0:properties}

\begin{figure}
  \includegraphics[width=\columnwidth]{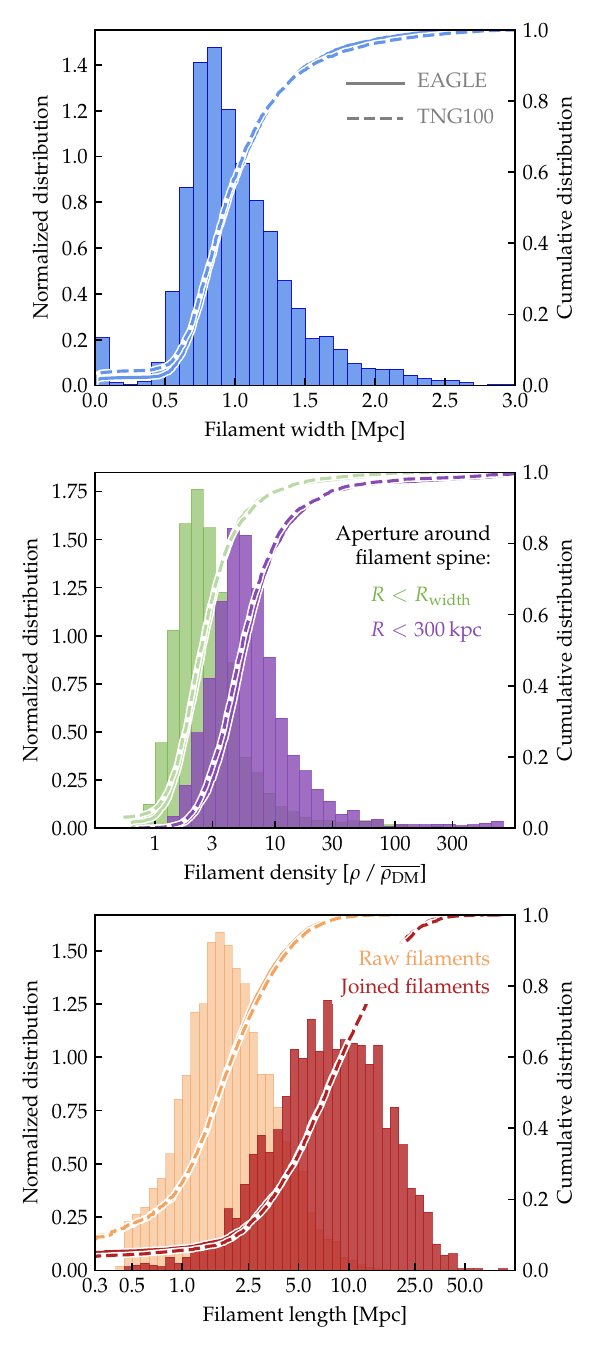}
  \caption{Filament properties at $z = 0$ for EAGLE and TNG100. \emph{Top panel:} filament widths, defined as described in the text and illustrated in Fig.~\ref{fig:6_filament_density_profiles}. \emph{Middle panel:} Average DM density within the filament width (green) and the central 300 kpc from the spine (purple). \emph{Bottom panel:} Lengths of the `raw' filaments as identified by DisPerSE (light red) and the combined filaments after our joining step (dark red; see Sec.~\ref{ssec:joining}). In all cases, histograms represent the differential distribution for EAGLE, while the solid and dashed lines show the cumulative distributions for EAGLE and TNG100, respectively. The two simulations agree almost perfectly, with typical values of $\sim$1 Mpc for width, $\sim$10$\times$ overdensity within 300 kpc, and $\sim$10 Mpc length for the joined filaments.}
  \label{fig:7_filament_properties}
\end{figure}

The full distribution of filament widths is displayed in the top panel of Fig.~\ref{fig:7_filament_properties}; solid and dashed lines show the cumulative distribution for EAGLE-Ref100 and TNG100, respectively, while the histogram gives the (normalized) differential distribution for EAGLE. It is evident that EAGLE and TNG100 agree extremely closely: the dashed and solid lines lie almost exactly on top of each other in all panels. The median width is 0.9 Mpc, with tails to 0.4 and 2.9 Mpc, respectively. These widths are broadly consistent with the distribution of \citet[their fig.~38]{Cautun2014}, and with the $\approx\,$1 Mpc width inferred from the stacked galaxy profile by \citet{Wang2024}, keeping in mind the different definitions. The dearth of filaments with a width of only a few 100 kpc may be an artefact from slightly inaccurately placed spines, but it could also be a (physical) manifestation of which filaments we have deemed `significant' in the calibration of the persistence threshold. A few per cent of filaments in both EAGLE and TNG100 have a central overdensity of $<\,$2, and hence a width of zero.

In the middle panel of Fig.~\ref{fig:7_filament_properties} we show the distribution of two measures of filament density: first (in green) the average out to the width of the filament (i.e.~where the density reaches twice the local background), and secondly (in purple) the average in the filament centre, out to 300 kpc from the spine. The latter is, unsurprisingly, higher, with median overdensity of 2.5 and 5, respectively; similar to the filament width both have a moderately broad distribution within a factor of $\sim$30 (2--60 for the central density, albeit with a tail to values well above 100 to which we will return below).

Finally, the bottom panel of Fig.~\ref{fig:7_filament_properties} shows the filament lengths, for both the raw (light red) and joined (dark red) filaments. The former has a median of 1.7 Mpc, while joining typically extends the length by a factor of $\approx\,$4, to a median of 7.3 Mpc. The full distribution extends over a factor 100, from as short as 0.5 Mpc for the shortest, to $\approx\,$50 Mpc for the longest (joined) filaments for both EAGLE and TNG100. This distribution of lengths is similar to what \citet{Galarraga-Espinosa2020} report for the lower-resolution IllustrisTNG300 simulation (their fig.~3), with filaments also identified by DisPerSE but using galaxies as tracers. In contrast, \citet{Cautun2014} find a $z = 0$ filament length distribution that is almost flat up to $\approx\,$20 Mpc and then extends to lengths beyond 100 Mpc. While the absence of such very long filaments in our work may be due the relatively small box size of EAGLE and TNG100, the different distributions at shorter lengths more likely arise from our different definitions of the filament network and its partition into individual filaments.

\begin{figure}
  \includegraphics[width=\columnwidth]{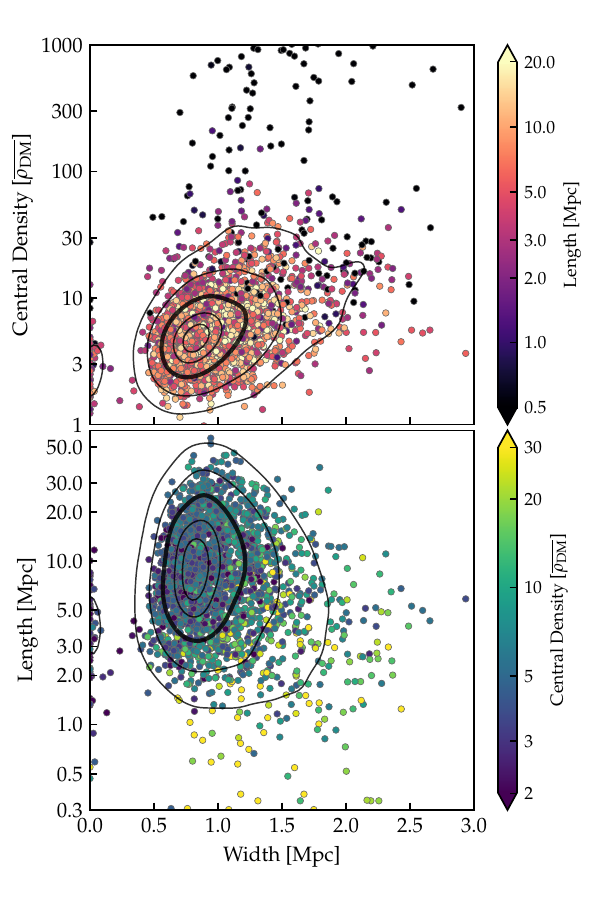}
  \caption{Correlation between filament width, length, and central density (within 300 kpc from the spine) for EAGLE at $z = 0$. The \emph{top panel} shows density vs.~width coloured by length, the \emph{bottom panel} length vs.~width coloured by density. For both, points represent individual filaments while black lines enclose the central 10, 25, 50 (thick), 75, and 90 per cent of points. There is a modest correlation between density and length, while length and width are largely independent apart from outliers with density $\gtrsim 50 \times$ the cosmic mean.}
  \label{fig:8_filament_property_correlations}
\end{figure}

The correlations between these three filament properties are shown in Fig.~\ref{fig:8_filament_property_correlations}; given the close agreement between the distributions of EAGLE and TNG100 in Fig.~\ref{fig:7_filament_properties} we only show EAGLE here. From the top panel, we see a moderately clear correlation between width and central density, in the sense that the thickest filaments also tend to be the densest, albeit with considerable scatter (an order of magnitude). The tail of filaments with extremely high central densities, $> 30 \overline{\rho_\mathrm{DM}}$ are an exception; they tend to have moderate widths ($\sim$ 1 Mpc) but---almost uniformly---extremely short lengths of $< 1$ Mpc. Visual inspection (not shown) confirms that they lie in the vicinity of reasonably massive haloes and are therefore arguably an artefact of our filament finding approach rather than genuine filaments.

Between length and width there is hardly any correlation (bottom panel of Fig.~\ref{fig:8_filament_property_correlations}), apart from a tendency for the shortest filaments to also be the thickest. This is very different from \citet{Cautun2014}, who find a clear positive correlation between filament length and width (their fig.~57), albeit with different definitions of filaments, their lengths, and widths from us. As discussed above, many of our short filaments instead have high central densities (yellow colour in the plot), although there are also a lot of `thick-and-short' filaments that only reach central overdensities of $\sim$10. Our analysis therefore suggests that cosmic filaments are quite heterogeneous and cannot be meaningfully classified by a single number. This is notably different from haloes, whose mass and concentration are strongly correlated so that the former provides a useful one-parameter label, at least to first order.

\subsection{Density profiles for different components}
\label{sec:z0:multiprofiles}

\begin{figure}
  \includegraphics[width=\columnwidth]{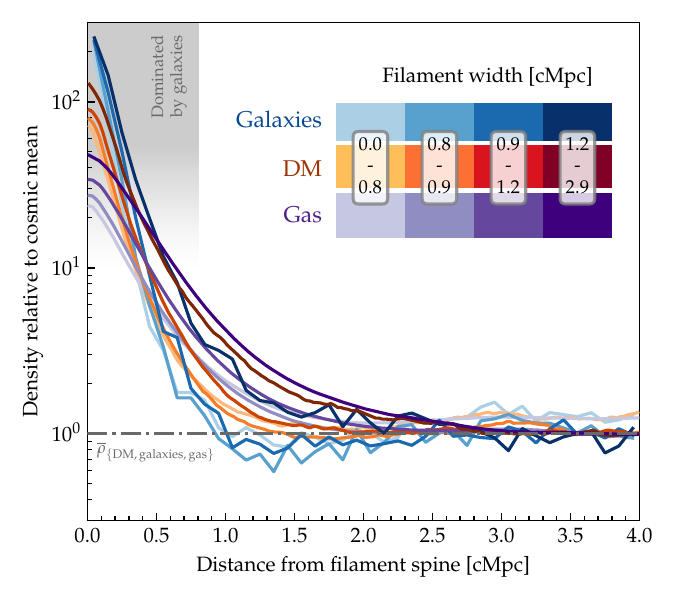}
  \caption{Mean-averaged density profiles of galaxies, DM, and gas around filaments in EAGLE at $z = 0$. Different shades represent different filament width quartiles as indicated. All profiles are mean-averaged across the relevant filament segments and normalised to their respective cosmic mean (grey dash-dotted line). The different profiles agree broadly with each other, with the galaxies most concentrated near the filament spine while the gas is most extended. Even in the thickest quartile, the average profiles reach the cosmic mean at $< 3$ Mpc from the spine.}
  \label{fig:9_filament_density_profiles}
\end{figure}

We have found above that the DM distribution within filaments is rather narrow and extends at most a few Mpc from the filament spine. In Fig.~\ref{fig:9_filament_density_profiles}, we compare this distribution to that of the baryonic components; galaxies, DM, and gas are shown in shades of blue, red, and purple, respectively. For clarity, we only show EAGLE here, but the profiles from TNG100 agree closely (see Fig.~\ref{fig:B1_filament_density_profiles_eagle_tng_z0} in Appendix \ref{app:density_profiles}). Darker shades represent thicker filaments, where the thickness is determined as described above, i.e.~from the DM. In contrast to Fig.~\ref{fig:6_filament_density_profiles}, we show here the mean, rather than the median, across segments (again weighted by segment length). This is necessary because the low galaxy number density does not permit us to construct meaningful galaxy density profiles for individual segments. For a direct comparison between the three components, we normalise each to their respective cosmic mean ($\overline{\rho}$).

The DM profiles (red shades in Fig.~\ref{fig:9_filament_density_profiles}) resemble those shown in Fig.~\ref{fig:6_filament_density_profiles}, albeit with a significantly higher and more cuspy central overdensity ($\sim$100 instead of $\sim$10) and an asymptote very close to, rather than below, the cosmic mean at $R \gtrsim 3$ Mpc. Both differences are due to the higher influence of segments with disproportionally high density on the mean compared to the median (we will return to this poin in Sect.~\ref{sec:gas}). In other words, the outskirts of filaments are slightly underdense over most of their volume (Fig.~\ref{fig:6_filament_density_profiles}), but contain almost exactly the average amount of DM once the contribution of local overdensities is taken into account.

Compared to the DM, the galaxy profiles are more strongly concentrated, with a central overdensity of $\sim$200 and a drop to the cosmic mean by $R = 1$ Mpc except for the thickest filament bin (dark blue). The stronger concentration of galaxies than DM towards the spine may partly be by construction (DisPerSE will tend to place filament spines close to overdensities, i.e.~galaxies), but as we argue in Sect.~\ref{sec:z0:substructure} below, it is more likely the result of enhanced halo formation near filament spines. Our galaxy profiles can be compared to \citet[their fig.~6]{Galarraga-Espinosa2020}. They find a similar overdensity close to the spine, but a much slower decline, with their profiles only reaching the cosmic mean at $\approx\,$20 Mpc. As we will show in Sect.~\ref{sec:galaxy-filaments}, this difference arises because their choice of using galaxies as filament tracers has systematic shortcomings.

Meanwhile, the gas displays the opposite behaviour of galaxies and is somewhat less strongly concentrated than the DM, with a central overdensity of $\sim$30 and a slightly more extended tail towards larger radii (by a few 100 kpc). The more extended profile of gas is, plausibly, the result of two physical effects: its non-zero pressure makes it more resistant to gravitational compression, while simulations also predict that it is highly susceptible to feedback from galaxies pushing it away from the filament centre (we will return to this point in Sect.~\ref{sec:gas}). Both effects could in principle be compensated by gas cooling, but as discussed by \citet{Tuominen2021} the density and metallicity of gas in $z = 0$ filaments lead to cooling times longer than a Hubble time so that significant cooling is unlikely. Our profiles agree reasonably well with those of \citet[their fig.~15]{Tuominen2021}, also based on EAGLE but with a different filament definition (the Bisous algorithm applied to galaxies). We find similar central overdensities, but their profiles have a more pronounced density core in the central $\approx\,$1 Mpc and only drop to the cosmic mean at a radius of $\approx\,$4 Mpc. Again, this is most likely a consequence of their use of galaxies as filament tracers, in addition to the intrinsic differences between Bisous and DisPerSE in the definition of filaments and their spines.

For all three components, there is a small but clear offset between the four bins in filament width, with the thickest filaments consistently having the highest density, at least out to $\approx\,$2.5 Mpc. At larger radii, there is a small but noteworthy inversion, with the thinnest filaments slightly overdense in all three components (by a factor $< 1.5$) while the thickest ones stay close to the cosmic mean. We see no such hint in Fig.~\ref{fig:6_filament_density_profiles}, so that this could be another subtle effect of mean-averaging. Alternatively, it might be a consequence of hierarchically correlated large-scale structure: thick filaments tend to be surrounded by thinner ones (see Fig.~\ref{fig:2_overview}), so that the outermost profiles of thin filaments may start to pick up some of the matter from their prominent neighbours. In any case, the increase is small and the average profile of all filaments remains within a factor 1.5 of the cosmic mean at $R \gtrsim 2$ Mpc. At very large distances ($R \gtrsim 15$ Mpc, not shown in Fig.~\ref{fig:9_filament_density_profiles}) we have found that all four width quartiles show a slight \emph{underdensity} of galaxies (by a factor $\gtrsim$0.8), because the profiles then preferentially probe the voids surrounding filaments.

As a final caveat, we note that our discussion above has implicitly assumed that filaments are well-described by cylindrical symmetry. While this may be appropriate on scales up to a few Mpc, it is worth keeping in mind that filaments are typically embedded within sheets so that the density on their outskirts is more complex, both due to the anisotropic distribution of diffuse matter within the sheets and the fact that neighbouring filaments will be preferentially located along the same direction. 

\subsection{How smooth are filaments?}
\label{sec:z0:substructure}

\begin{figure}
  \includegraphics[width=\columnwidth]{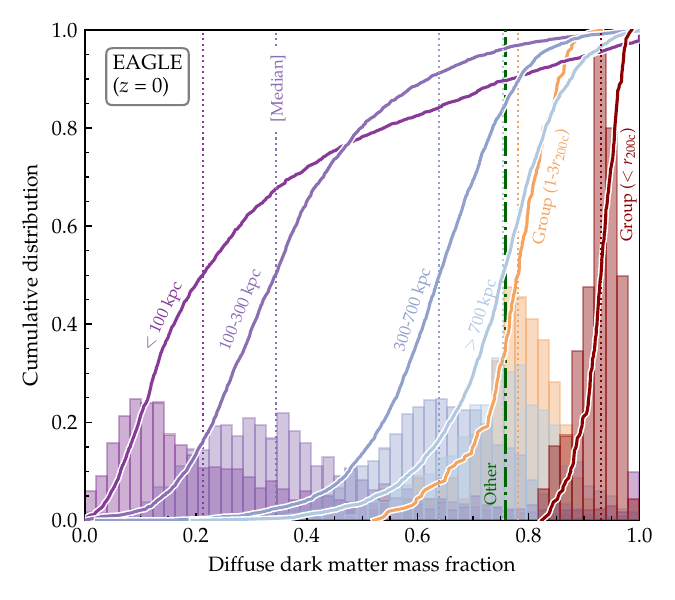}
  \caption{Diffuse dark matter mass fraction $f_\mathrm{d}$ in filaments. Different shades of purple/blue represent different radial zones around the filament spines. Histograms and solid lines show the differential and cumulative distributions of diffuse fractions over individual filaments, with the correspondingly coloured dotted vertical line giving the median fraction. The analogous distributions for the centre and outskirts of groups are shown in red and orange, respectively, while the dash-dotted green vertical line marks the average diffuse fraction of dark matter outside of groups or identified filaments. Most filaments are substructure dominated in their centre (low $f_\mathrm{d}$), in contrast to groups.}
  \label{fig:10_filament_smoothness}
\end{figure}

We noted above that DM filaments contain prominent substructures (Fig.~\ref{fig:1_filament_close_up}), which significantly affect density profiles near the filament spine (Fig.~\ref{fig:9_filament_density_profiles}). To understand their role more clearly, it is instructive to quantify the substructure fraction within filaments (we focus on EAGLE for simplicity, but expect identical results for TNG100). We separate all DM particles in the simulation into those belonging to individual filaments (closer to a spine than the width of the filament); individual groups (within $3\, r_\mathrm{200c}$ of a halo with $M_\mathrm{200c} > 10^{13}\,\msun$); and others, which include particles in voids, walls, and in filaments below our detection threshold. The filament particles are further subdivided into four radial bins around the spines (0--100 kpc; 100-300 kpc; 300-700 kpc; and $>$700 kpc), while for groups we distinguish between particles within and beyond $r_\mathrm{200c}$. For each resulting set of DM particles, we then identify those that are not gravitationally bound to a resolved subhalo (excluding the central subhalo for groups), i.e.~the diffuse `background' of their environment rather than a clumpy substructure. The resulting distributions of diffuse mass fractions $f_\mathrm{d}$ are shown in Fig.~\ref{fig:10_filament_smoothness}.

There is a very clear difference between filaments and groups: while the diffuse background is dominant for the latter (median $f_\mathrm{d} = 0.93$ within $r_\mathrm{200c}$), its mass fraction is much lower in filaments, with a strong trend to lower $f_\mathrm{d}$ closer to the spine (a median of only 0.21 within 100 kpc). Remarkably, the outskirts of groups and filaments (light blue and orange, respectively) have almost the same diffuse DM fractions, with a median close to that outside filaments or groups ($f_\mathrm{d} \approx 0.75$). In other words, groups (and clusters) are predominantly smooth structures surrounded by average-clumpy outskirts, while filaments have a highly clumped core surrounded by a comparatively smooth outer region. As we will show in Paper II, similar differences are seen for the halo mass function.

A second noteworthy difference between filaments and groups is the object-to-object scatter in $f_\mathrm{d}$, with groups having uniformly high values (only 10 per cent below 0.88) while the central filament bin spans almost the full range from $f_\mathrm{d} = 0$ to 1. We have found no significant trend of $f_{\mathrm{d},\,\leq 100\,\mathrm{ kpc}}$ with filament length or width (not shown), but there is a clear correlation with total mass in the same aperture, with a median of 0.70 and 0.10 for the least and most massive quartile, respectively\footnote{The correlation with mass per unit length is similarly strong (medians of 0.66 and 0.08 for the least and most massive quartile, respectively), whereas that for only the diffuse mass is weaker but still clear (0.55 vs.~0.14). The latter rules out that the trend is self-induced by the presence of the substructures.}. 

The high diffuse fraction in groups is easily understood as arising from the stripping and disruption of (sub)haloes in the strong tidal field near the group centre (e.g.~\citealt{Hayashi2003,Bahe2019}). In filament cores, another mechanism must be dominant. Discounting the possibility that the high substructure fractions are the cause and not the effect of their proximity to the filament spines---which would be inconsistent with the good agreement between the spines and the DM field as seen in Figs.~\ref{fig:2_overview}, \ref{fig:4_disperse_variations}, and \ref{fig:5_joined_filaments} (and also in Figs.~\ref{fig:17_single_filament_z0} and \ref{fig:18_single_filament_z2} below)---there are two plausible explanations: accretion of dark matter onto haloes could be more efficient in filament cores, or haloes might preferentially accumulate near filament spines. Both of these are physically plausible and consistent with the higher substructure fraction in more massive filament cores: the overdensity of filaments corresponds to a higher volume density of haloes, so that each DM particle is closer to a halo and hence more easily accreted, while dynamical friction provides a natural mechanism for the segregation of haloes towards the local gravitational potential minimum along filament spines. Most likely, both are at play to some extent, but quantifying their respective roles would be a non-trivial task beyond the scope of this work.

\section{Filaments at redshift $z = 2$}
\label{sec:z2}

\subsection{The cosmic web at $z \gg 0$}

We have so far focused our analysis on the present-day Universe, where the formation of large-scale structure has progressed the furthest and the observational picture is clearest. Nevertheless, it is now clear that the inhomogeneous nature of the Universe is not only a low-redshift phenomenon: galaxy clusters have been identified up to $z \approx 2$ (e.g.~\citealt{Nantais2016, Balogh2017, Belli2019, Xu_C_2023, Calzadilla2023}) and overdensities that may correspond to proto-clusters up to $z \approx 8$ (e.g.~\citealt{Laporte2022,Morishita2023,Champagne2024}). What is more, while systematic explorations of filaments in three dimensions are currently limited to $z \lesssim 1$, there are already tentative detections of individual filament-like structures out to $z \approx 4$ (e.g.~\citealt{Tornotti2024, Tornotti2024a}) and deep spectroscopic surveys such as MOONRISE \citep{Maiolino2020} are about to push the redshift limit of systematic filament studies beyond $z = 2$. It is therefore of considerable interest to test in how far high-redshift filaments resemble those found in the local Universe.

\begin{figure*}
  \includegraphics[width=\textwidth]{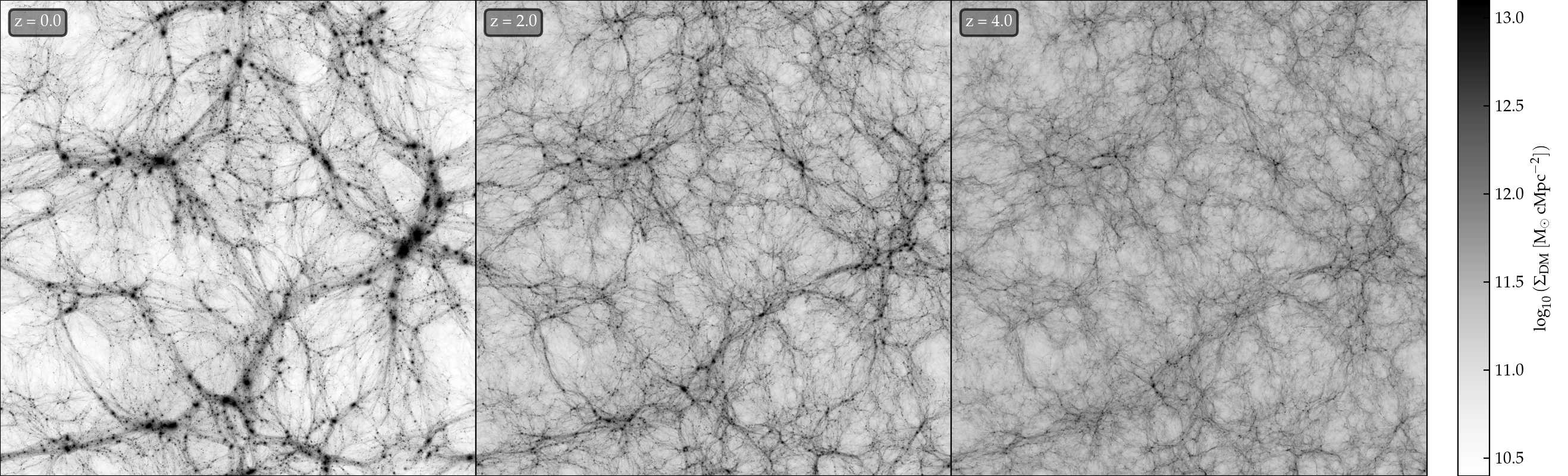} 
  \caption{Cosmic web structure in EAGLE-Ref100 at redshifts 0, 2, and 4 (left, middle, and right panels, respectively). All three panels show the DM density projected over the same 15 cMpc thick slice, with the same density scaling in co-moving units. Going back in time, the cosmic web is very clear at $z = 0$, still recognizable at $z = 2$, but almost dissolved by $z = 4$.}
  \label{fig:11_cosmic_web_evolution}
\end{figure*}

As a first step, we show  in Fig.~\ref{fig:11_cosmic_web_evolution} a visual representation of the cosmic web at redshifts $z = 0$, $2$, and $4$. Each panel displays the projected DM density of EAGLE-Ref100 within the same $100\, \times\, 100\, \times\, 15$ cMpc slice. We use the same density scaling (in co-moving units) for all three panels, so that we recover the expected trend of a significantly decreasing density contrast towards higher redshift, driven in particular by the less developed underdensities in voids. Nevertheless, a clear filamentary structure spanning the full panel remains visible at $z = 2$, albeit with generally thinner filaments and without the prominent cluster-scale haloes at the end of the thickest filaments like at $z = 0$. At a qualitative level, it is evident that the $z = 2$ filaments already occupy broadly the same regions\footnote{We note that the simulation employs periodic boundary conditions, so that the left and right edges correspond to the same physical region. This is why many of the prominent filaments continue fairly smoothly across adjacent panels.} that will host the major filaments and nodes at $z = 0$. This general picture of the cosmic web at $z > 0$ agrees with previous works (e.g.~\citealt{Gheller2016,Martizzi2019}) and is also consistent with the result of \citet{Cautun2014} that $\approx\,$90 per cent of the mass in $z = 2$ filaments will end up in filaments or nodes at $z = 0$.

At $z = 4$, the picture looks very different. There are hardly any moderately prominent haloes left, and the remaining large-scale structure consists of extremely thin `proto-filaments', which often appear bundled together but are relatively short individually and form less of a coherent, percolating network. It is unclear which of these structures we would want to identify as filaments for a meaningful comparison to lower redshifts, even leaving aside the technical question of how well a filament finder would perform in this regime. As a result of all these considerations, we focus our quantitative high-redshift analysis on $z = 2$.

\subsection{Filament identification at $z = 2$}

\begin{figure*}
  \includegraphics[width=\textwidth]{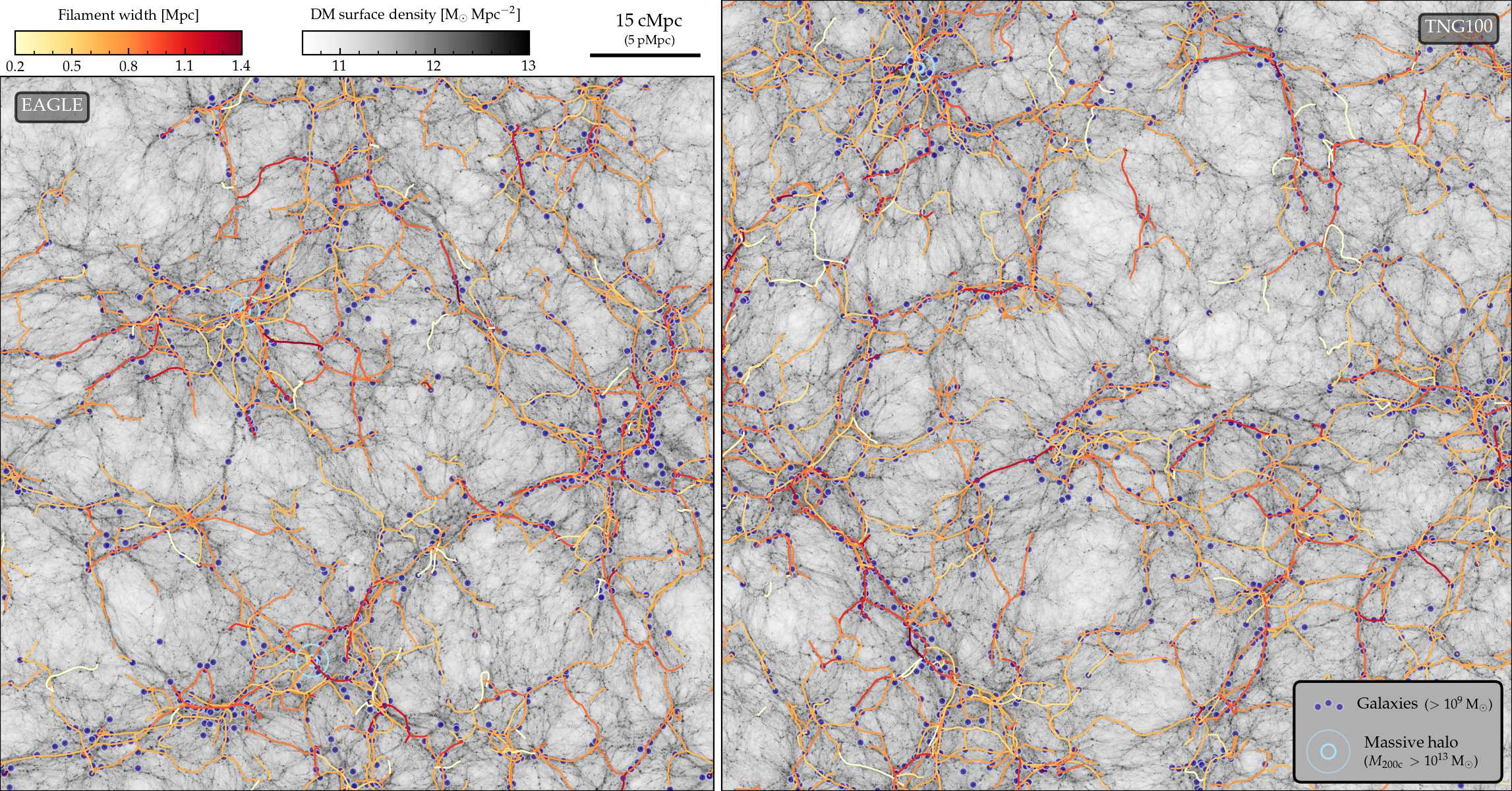} 
  \caption{Filamentary skeleton at $z = 2$, analogous to the $z = 0$ skeleton shown in Fig.~\ref{fig:2_overview}. EAGLE is shown on the left, TNG100 on the right. Red/orange lines represent filaments in a 15 cMpc thick slice, background images show the DM density projected in the same slice. Galaxies with $M_\mathrm{star}\,(z = 2)\, > 10^9\,\mathrm{M}_\odot$ are shown as indigo circles. As for $z = 0$, the filaments identified by DisPerSE correspond well to the main features of the DM density field, complemented by a myriad of unidentified thinner filaments.}
  \label{fig:12_skeleton_z2}
\end{figure*}

We identify filaments at $z = 2$ with nearly the same approach as detailed for $z = 0$ in Sect.~\ref{sec:filaments:disperse}. The only difference is that we have re-calibrated the DisPerSE persistence threshold, and use a 50 per cent higher value (i.e.~$6 \cdot 10^7$ instead of $4 \cdot 10^7\,\msun / V_\mathrm{cell}$). Our reasoning for this change is analogous to the calibration at $z = 0$, namely that a value of $4 \cdot 10^7$ would have led to a too dense filamentary network that would only have covered 7 per cent more galaxies (92 vs.~85 per cent of all galaxies with $\mstar > 10^9\,\msun$, within 1 cMpc from the spines) but with a 64 per cent greater total length. Our resulting fiducial filament network at $z = 2$ is shown in Fig.~\ref{fig:12_skeleton_z2} for both EAGLE and TNG100, in analogy to Fig.~\ref{fig:2_overview} for $z = 0$.

As can be seen, our DisPerSE approach results in a good representation of the DM filaments at $z = 2$, much the same as at $z = 0$ (Fig.~\ref{fig:2_overview}). This confirms that we can meaningfully analyse the filaments at this redshift, and compare the result to $z = 0$. As already noted above, a profound difference is the near-absence of massive haloes ($M_\mathrm{200c} > 10^{13}\,\msun$), of which there are only 2 and 1 in the displayed $z = 2$ slice for EAGLE and TNG100, respectively (in the full EAGLE and TNG100 volumes, there are 18 and 25 such haloes, respectively, as opposed to 162 and 182 at $z = 0$). Instead, the locations of many of the prominent haloes at $z = 0$ are now the sites of thick filaments arranged in a web-like pattern, compared to the predominantly radial orientation of filaments around the massive haloes at $z = 0$. As at $z = 0$, the void regions between our identified filaments contain almost no galaxies but are spanned by a large number of tenuous DM filaments that fall below our DisPerSE detection threshold.

\subsection{Filament properties at $z = 2$}

\begin{figure}
  \includegraphics[width=\columnwidth]{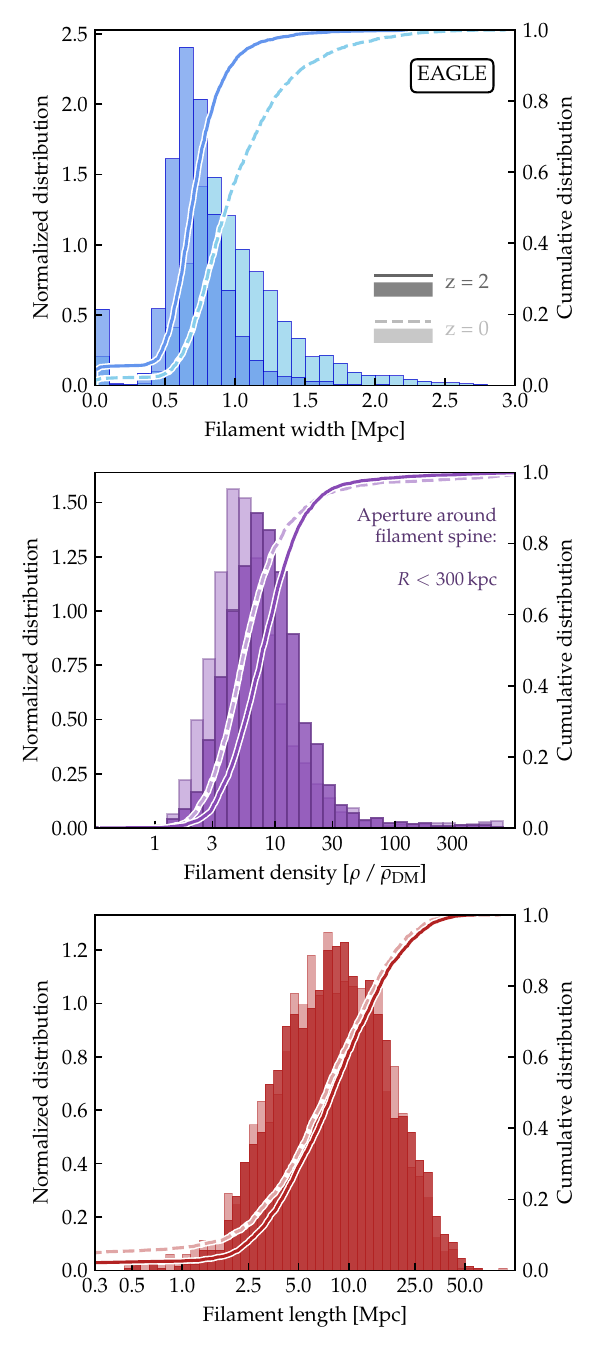}
  \caption{Filament properties at $z = 2$ (dark colours and solid lines) compared to $z = 0$ (light colours, dashed lines). For clarity, only EAGLE is shown. In analogy to Fig.~\ref{fig:7_filament_properties}, we show the filament widths, central densities, and lengths in the top, middle, and bottom panels, respectively. Histograms represent the (normalized) differential distribution, whereas lines give the corresponding cumulative distributions. Even in co-moving coordinates, filaments are $\approx\,$30 per cent thinner at $z = 2$ than at $z = 0$, while their central overdensities and (comoving) lengths are consistent between the two epochs.}
  \label{fig:13_filament_properties_z2}
\end{figure}

With the $z = 2$ filaments identified, we determine their width, density, and length in the same fashion as for $z = 0$. For the width and length, one subtlety is that at $z > 0$ we have the choice between proper and co-moving units. We choose the latter, because filaments remain significantly coupled to the cosmic expansion, as is evident from Fig.~\ref{fig:11_cosmic_web_evolution}. With this choice, the width of $z = 2$ filaments as shown in dark blue in the top panel of Fig.~\ref{fig:13_filament_properties_z2} is similar to, but generally smaller than, that at $z = 0$ (light blue histogram and dashed cumulative distribution). The median width is 0.7 cMpc (vs.~0.9 cMpc at $z = 0$), with only 5 per cent of $z = 2$ filaments wider than 1.2 cMpc (vs.~35 per cent at $z = 0$). For clarity, we only show the results for EAGLE here but have verified that, as for $z = 0$, the equivalent distributions for TNG100 agree closely.

The global distribution of central overdensities and filament lengths are not very different between $z = 2$ and $z = 0$ (middle and bottom panels of Fig.~\ref{fig:13_filament_properties_z2}): both are shifted to slightly higher values, but the offset is negligible compared to the spread of values at either redshift. These small shifts might, conceivably, be the result of our slightly higher persistence threshold at $z = 2$ (see above), although one would then also expect a shift towards thicker, rather than thinner, filaments. In light of this uncertainty, we caution against over-interpreting this result and conclude instead that the (central) overdensity and comoving length of filaments have, on average, remained almost unchanged over the past 10 Gyr. It is worth keeping in mind that we are not tracking individual filaments, but compare populations that are only partly progenitors and descendants of each other: not all $z = 0$ filaments will have identified progenitors at $z = 2$, while some of the $z = 2$ filaments may have fully collapsed into nodes by $z = 0$ (see e.g.~\citealt{Cautun2014} and \citealt{Galarraga-Espinosa2024}).

We have also constructed the mass-weighted DM, gas, and galaxy profiles around $z = 2$ filaments in analogy to those shown in Fig.~\ref{fig:9_filament_density_profiles} for $z = 0$. There are no qualitative differences compared to $z = 0$, so that we only show these in Appendix \ref{app:density_profiles} (Fig.~\ref{fig:B2_filament_density_profiles_eagle_z2}). Quantitatively, the central mass-weighted DM densities are slightly lower than at $z = 0$ (reaching an overdensity of $\approx\,$50 instead of $\approx\,$100), so that the difference between DM and gas is less pronounced\footnote{Above, we had found that the central density from the length-weighted median profiles are if anything slightly higher at $z = 2$ than at $z = 0$. This difference is due to the higher sensitivity of mean-averaged profiles to embedded haloes, and their lower abundance at higher redshift.}. Likewise, the diffuse DM fractions $f_\mathrm{d}$ at $z = 2$ are qualitatively similar to $z = 0$ (see Sect.~\ref{sec:z0:substructure}), although the median $f_\mathrm{d}$ in filament cores is slightly higher (0.34 vs.~0.21). Consistent with our visual impression from Fig.~\ref{fig:11_cosmic_web_evolution}, this means that filaments are somewhat less clumpy at higher redshift, although still far less smooth than group or cluster haloes.

\subsection{Comparison to other works}

There are a number of other studies that have investigated the evolution of filament lengths, widths, and/or densities in cosmological simulations \citep{Cautun2014,Zhu2021,Galarraga-Espinosa2024,Wang2024}. While differences in the precise definition of these quantities make it difficult to compare results in detail, we nevertheless attempt an approximate comparison with these works.

In terms of filament widths, our finding of a shift towards slightly thinner filaments at $z = 2$ is broadly consistent with \citet{Zhu2021}, who also report a deficiency of the thickest filaments, and a corresponding increase in the thinnest ones, towards higher redshift (their fig.~2). \citet{Cautun2014} find no such shift in the width distribution (their fig.~38), although like \citet{Zhu2021} they define filaments based on the tidal tensor of the DM field. Likewise, \citet{Wang2024} find little change in the (stacked) filament width from $z = 0$ to 2, based on DisPerSE but with galaxies as tracers of filaments and profiles. Towards higher redshifts, they report a strong thickening of filaments, by a factor $\approx\,$2.5 from $z = 2$ to 4. We do not quantify our filament widths at $z > 2$, but it is evident from Fig.~\ref{fig:11_cosmic_web_evolution} that DM filaments become, if anything, even thinner at $z = 4$ than at $z = 2$. \citet{Galarraga-Espinosa2024} do not quantify the filament width explicitly, but show that their radial galaxy overdensity profiles are almost unchanged from $z = 0$ to 4 (their fig.~13). Most likely, the differences between \citet{Wang2024}, \citet{Galarraga-Espinosa2024}, and our work are due to the different definitions of filament width, as well as their use of galaxies as filament tracers (see Sect.~\ref{sec:galaxy-filaments}).

We are not aware of another study that has computed the DM density in the core of individual filaments. However, our result of a nearly unchanged density distribution between $z = 0$ and 2 agrees qualitatively with \citet{Cautun2014}, who found similarly small differences between the full DM density distributions in filaments at $z = 0$ and 2 (their fig.~26); and (indirectly) with the near-constant galaxy density profile of \citet{Galarraga-Espinosa2024}.

For filament lengths, \citet{Cautun2014} report a noticeable shift towards shorter filaments at $z = 2$, with filaments of length $\approx\,$7 ($\approx\,$60) cMpc three times more (less) abundant than at $z = 0$ (their fig.~53). In contrast, the length distributions shown by \citet{Galarraga-Espinosa2024} are almost redshift-invariant, with only a $\approx\,$10 per cent shift towards larger values. Our redshift-invariant length distribution is in closer agreement with their result than with \citet{Cautun2014}. Since the filaments of \citet{Galarraga-Espinosa2024} are, like ours, based on DisPerSE (albeit with galaxies as tracers), this demonstrates once again that alternative filament definitions can lead to significant differences in their properties.

\section{Gas in and around filaments}
\label{sec:gas}

Since the gas component within filaments is expected to be most closely connected to the evolution of galaxies, and its baryonic nature gives rise to more complex phenomena than DM, its properties merit closer investigation. We first analyse the gas temperature profiles (Sec.~\ref{sec:gas:tprofiles}) and then inspect maps of gas density, temperature, pressure, and metallicity (Sec.~\ref{sec:gas:maps}).

\subsection{Gas temperature profiles}
\label{sec:gas:tprofiles}

\begin{figure}
  \includegraphics[width=\columnwidth]{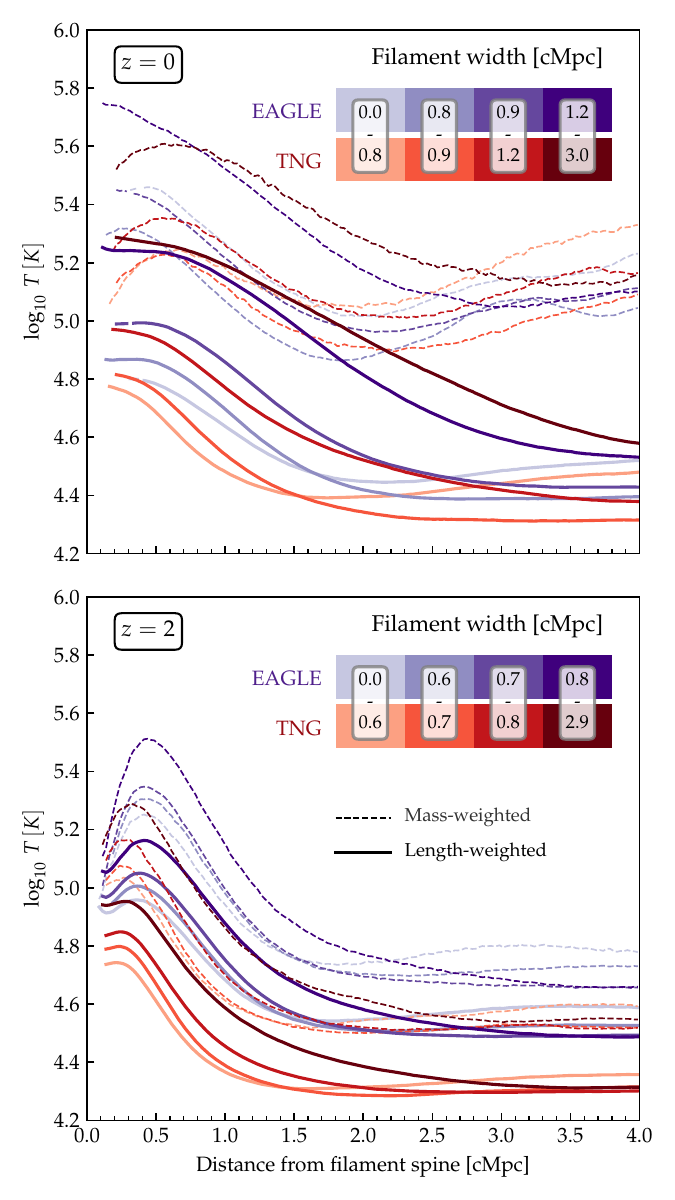}
  \caption{Gas temperature profiles around filament spines at $z = 0$ (top) and $z = 2$ (bottom), for the EAGLE and TNG simulations (shades of purple and red, respectively). Filaments are split by their width as indicated in each panel, with darker shades corresponding to thicker filaments. For each bin, we show both the mass-weighted profile (thin dashed lines) and the length-weighted one (thick solid lines); see text for details. While temperatures are generally decreasing towards larger radii, there is considerable variation with filament width, weighting method, and at $z = 2$ also between the two simulations.}
  \label{fig:14_temperature_profiles}
\end{figure}

To complement the insight from the gas density profiles in Fig.~\ref{fig:9_filament_density_profiles}, we analyse the gas temperature\footnote{Neither EAGLE nor TNG100 model the temperature structure of cold star-forming gas realistically, with the temperatures as used for the hydrodynamics calculation typically far exceeding the physical temperatures of the gas. We therefore assume a fixed value of $T = 10^4$ K for each gas particle or cell that has a non-zero star formation rate, but we have verified that this choice has no impact on our results shown here.} in Fig.~\ref{fig:14_temperature_profiles}. Very broadly, the typical gas temperature near the filament spines is several times higher than at 3--4 Mpc distance, in both simulations (red shaded lines for EAGLE, purple shades for TNG100) and at both redshifts ($z = 0$ in the top panel, $z = 2$ below). This is plausibly a consequence of accretion shocks and/or heating by feedback from galaxies embedded within the filaments, to which we return below.

In analogy to the density profiles, the presence of substructures and inhomogeneities within the filaments means that an average profile can only give a simplified description of the true filament structure. Different equally valid ways of averaging the temperature profile may therefore give substantially different results, so that we show results from two alternative approaches in Fig.~\ref{fig:14_temperature_profiles}. In each case, we average the (base-10) logarithm of the temperature, to reduce the impact of individual gas elements with extremely high temperature due to recent feedback events. Starting from the mass-weighted average temperature of all gas in each radial bin of each segment, we first weight all segments within a filament only by their length (solid lines), irrespective of the mass of gas that they contain at a given radius. This approach gives a representative temperature along the entire filament, although one that may not be characteristic of the bulk of its mass. As a second choice, we weight the segments by the mass of gas within each radial bin (thin dashed lines). In this way, segments that contain more gas at a given radius contribute more to the average, and we obtain a temperature that is characteristic of the majority of gas at this radius. The drawback is that this temperature may be biased by a small number of atypically dense segments and hence not necessarily representative of the filament along its full length.

\subsubsection{Temperature profiles at $z = 0$}

At $z = 0$ (top panel of Fig.~\ref{fig:14_temperature_profiles}), there is indeed a significant difference between the length- and mass-weighted profiles, the latter being systematically hotter by $\gtrsim 0.5$ dex. In other words, denser filament parts typically have a higher gas temperature. 
Focusing first on the length-weighted profiles (solid lines), we see that the thickest filaments (darkest colours) are systematically hotter at each radius, by up to a factor $\approx\,$3. The temperature profiles decrease smoothly with increasing distance from the spine and reach a floor of $T \approx 3\cdot10^4$ K at $R \approx 1.5$ Mpc for the thinnest filaments, but only at $R > 4$ Mpc for the thickest ones. EAGLE and TNG100 follow each other quite closely, although the latter (in red) predicts a marginally larger spread in temperatures across different filament thicknesses.

The mass-weighted $z = 0$ profiles (thin dashed lines) are not just offset from their length-weighted counterparts, but have qualitatively different features. For TNG100 (red shades), the central $\approx\,$0.5 Mpc show a clear downturn in all four bins of filament width, by 0.1--0.2 dex. For EAGLE, a (marginal) downturn is only seen for the two thinnest bins. In addition, all mass-weighted profiles show a minimum at $\approx\,$2--4 Mpc and rise again beyond this point. Both features suggest that gas in dense regions along filaments has distinct properties; we will return to this point below.

Compared to our $z = 0$ temperature profiles, the ones published by \citet[their fig.~5]{Galarraga-Espinosa2021}, \citet[their fig.~13]{Tuominen2021}, and \citet[their fig.~6]{Zhu2021} all show more pronounced isothermal cores near the spine, extending to $\approx\,$1--3 Mpc. As the study of \citet{Tuominen2021} is also based on EAGLE, and \citet{Galarraga-Espinosa2021} on the IllustrisTNG300 simulation that is closely related to TNG100, these differences are most likely due to different filament definitions: both of these studies choose galaxies as tracers, while we base the extraction of the cosmic web on the more detailed DM density field (see Sect.~\ref{sec:galaxy-filaments}).

Our temperatures are similar to \citet{Tuominen2021} in both the filament core and outskirts, but even our mass-weighted profiles have somewhat lower central temperatures than \citet{Galarraga-Espinosa2021}, and even more so than \citet{Zhu2021}, who like us find higher temperatures for thicker filaments. Differences in simulations and/or filament finder are plausible origins for these discrepancies.


\subsubsection{Temperature profiles at $z = 2$}

At $z = 2$ (bottom panel of Fig.~\ref{fig:14_temperature_profiles}), the situation is somewhat different. Although the maximum temperatures are similar to $z = 0$, broadly around $10^5\,\mathrm{K}$, the profiles drop much more rapidly and all reach a floor by $\approx\,$2 cMpc from the spine. Whereas the level of this floor was broadly consistent between EAGLE and TNG100 at $z = 0$, it is clearly different at $z = 2$, at $\log_{10} T\,\mathrm{[K]}\approx 4.3$ and 4.5 for TNG and EAGLE, respectively for the length-weighted profiles (solid lines). A similar offset is seen for the mass-weighted profiles, which (as for $z = 0$) are generally higher, by $\approx\,$0.2 dex. A shift to slightly lower temperatures at higher redshift has also been found by \citet[their fig.~7]{Zhu2021}. In contrast to us, they see an order-of-magnitude offset between different widths even at $z = 2$, which we attribute to our different definitions of filaments and their widths.

There is no pronounced rise in the gas temperatures at large radii at $z = 2$, but the drop in the innermost $\approx\,$500 ckpc is significantly stronger than at $z = 0$.  For EAGLE (purple lines), the mass-weighted profiles drop by $\approx\,$0.5 dex in the inner 0.5 Mpc, and even the length-weighted profiles (solid purple lines) have a $\approx\,$0.1--0.2 dex drop. The downturn is less pronounced for TNG, but its mass-weighted profiles still decrease by up to $\approx\,$0.2 dex in the central 300 kpc. Only the length-weighted TNG profiles show no central drop, just a plateau in the central $\approx\,$300 kpc. While it must be kept in mind that trends within the central $\approx\,$250 kpc may be affected by inaccuracies in the spine placement, such uncertainties should only act to wash out trends, not create a turnover and central dip. We are therefore confident that these trends are real and indicate the wide-spread presence of cooler gas close to the filament spine of $z = 2$ filaments.

\subsection{A detailed view of the gas structure in and around filaments from projected maps}
\label{sec:gas:maps}

To help interpret the temperature profiles discussed above, and gain further insight into the gaseous component of filaments, we inspect maps of key gas properties. Since we are here primarily interested in depicting the true structure of individual filaments, rather than sampling a significant fraction of the simulation volume, we integrate over thin slices to minimise confusion from projection effects\footnote{Since SPH describes a field in which gas properties can be calculated at any point, it would in principle be possible to eliminate projection effects completely and show a `razor-thin' slice. We do not do this here because no filament would lie perfectly along such a slice.}.

\begin{figure*}[t]
  \includegraphics[width=\textwidth]{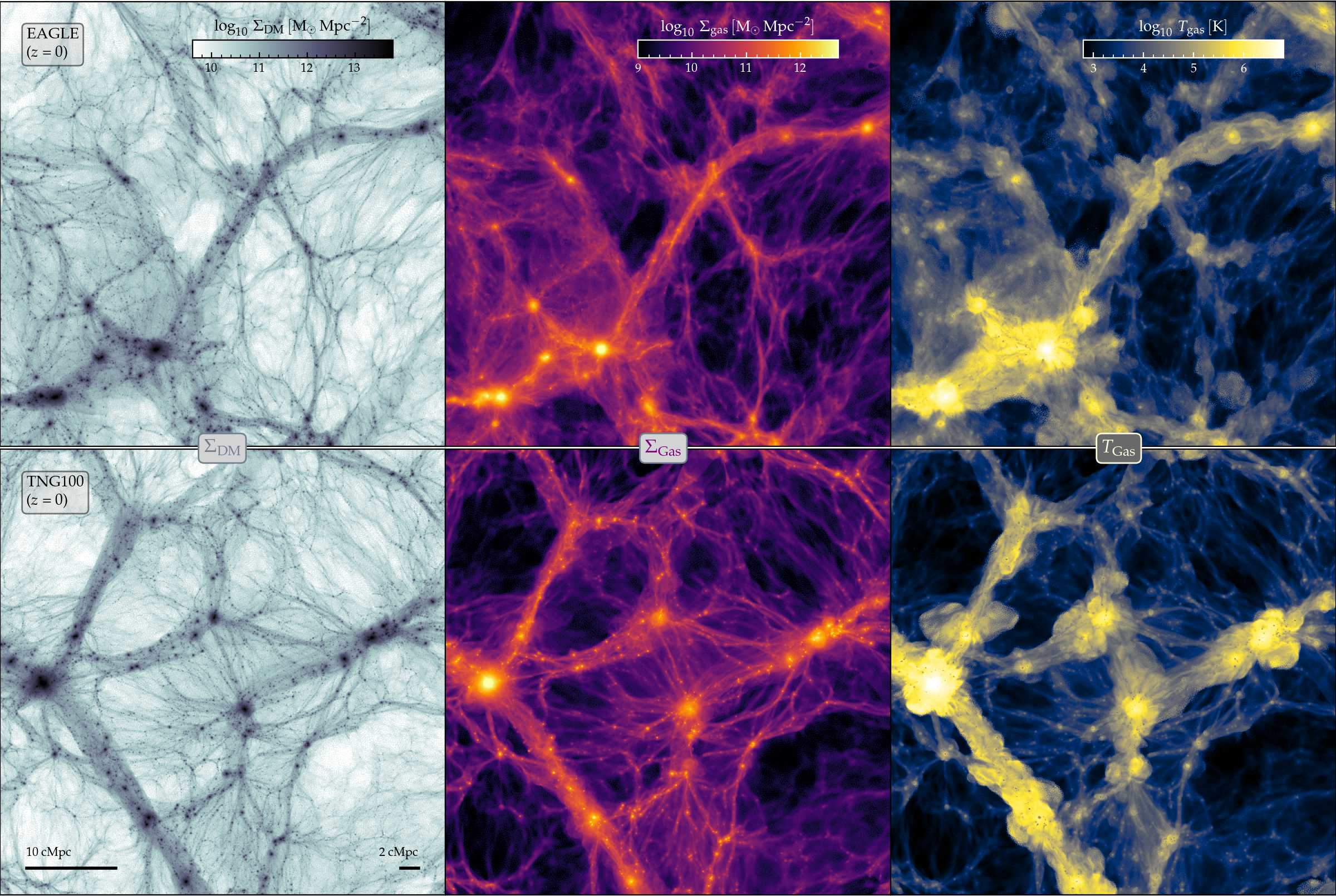}
  \caption{Structure of the DM and gaseous cosmic web at $z = 0$ in EAGLE (top row) and TNG100 (bottom row). All six panels show a field of view of $50 \times 50$ Mpc, with a projection depth of 5 Mpc. DM density is shown in the left-hand panels, gas density in the middle, and mass-weighted gas temperature on the right. While the DM structures are near-indistinguishable, there are clear differences between the simulations in both gas density and temperature, as discussed in the text. Particularly noteworthy is the prominence of small gas-rich haloes in TNG that are absent in EAGLE, and the ubiquity of warm gas around filaments in EAGLE.}
  \label{fig:15_cosmic_web_z0}
\end{figure*}

\begin{figure*}[t]
	\includegraphics[width=\textwidth]{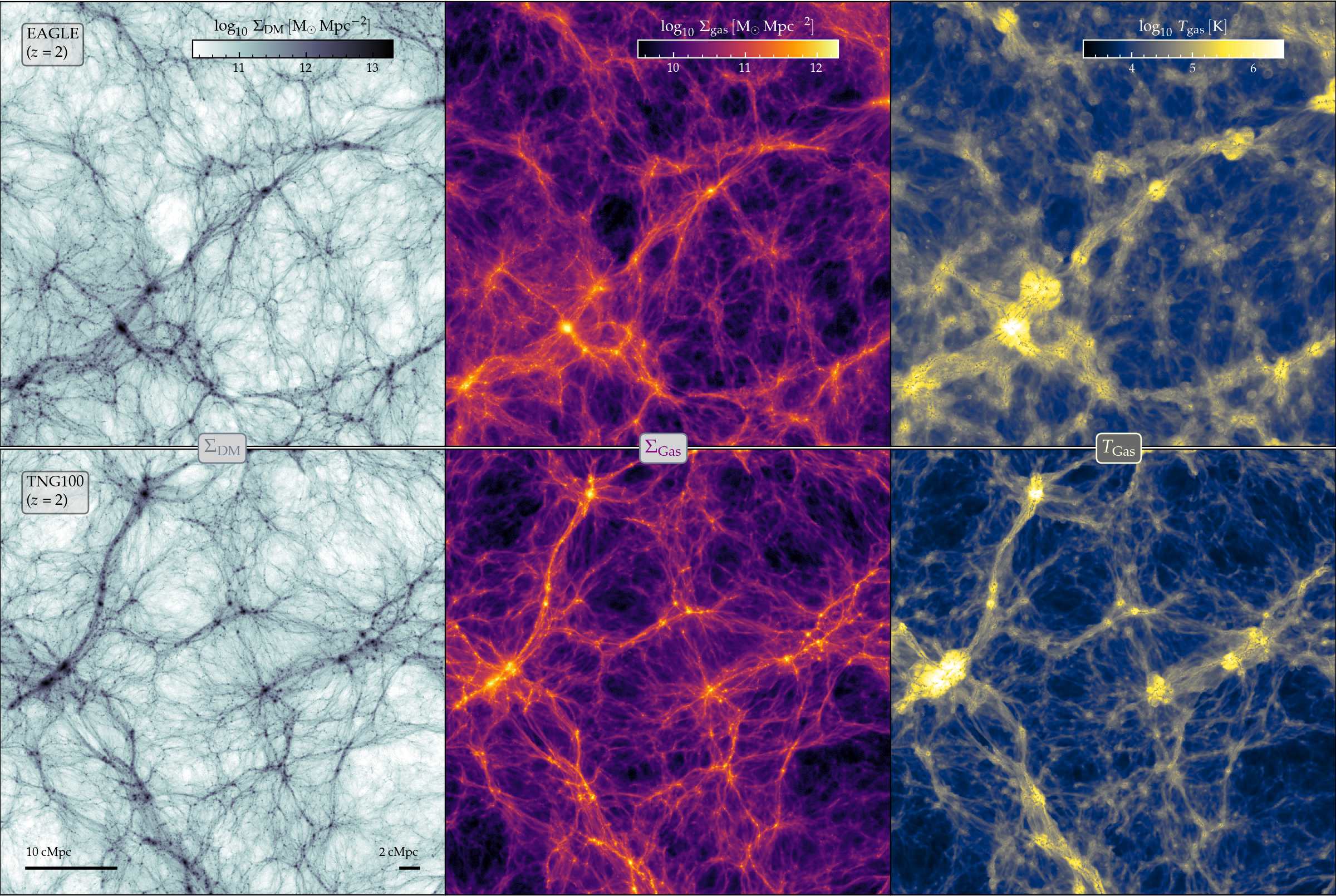}
	\caption{Structure of the DM and gaseous cosmic web at $z = 2$ in EAGLE (top row) and TNG100 (bottom row). The layout is analogous to Fig.~\ref{fig:15_cosmic_web_z0}; the six panels show the same $50 \times 50$ cMpc field, with a projection depth of 5 cMpc. DM density is shown in the left-hand panels, gas density in the middle, and gas temperature on the right. While the DM structure is consistent across both simulations, the gas shows a more diffuse structure in EAGLE with a hotter temperature across most of the field. In contrast to $z = 0$, both EAGLE and TNG100 clearly predict the existence of cold gas in the centre of the filaments, as discussed in the text.}
	\label{fig:16_cosmic_web_z2}
\end{figure*}

First,  Figs.~\ref{fig:15_cosmic_web_z0} ($z = 0$) and \ref{fig:16_cosmic_web_z2} ($z = 2$) show a $50 \times 50$ cMpc field with a projection depth of 5 cMpc, comparable to the diameter of filaments but still sampling a representative variety of structures. We display their gas surface density $\Sigma_\mathrm{gas}$ and mass-weighted gas temperature $T$ (averaging the logarithm of $T$, as for the profiles), directly comparing EAGLE (top rows) and TNG100 (bottom rows). To put these maps in context, we also show the corresponding surface density of DM in the left-hand columns. The DM maps of the two simulations resemble each other as closely as could be expected, given that they show two completely different regions of space. Any significant differences between the gas properties must therefore be due to baryon physics differences between the two simulations.

Haloes are much less prominent than in Figs.~\ref{fig:2_overview}, \ref{fig:11_cosmic_web_evolution}, or \ref{fig:12_skeleton_z2}. This is a consequence of the smaller projection depth, which decreases the probability for each line of sight to intersect a halo\footnote{Due to their rather high overdensities ($\gtrsim\,$100), haloes anywhere along the line of sight tend to dominate the surface density of a pixel. In contrast, the lower overdensities of filaments and walls ($\lesssim\,$10) tend to average out in projection and form a more or less uniform background. As a consequence, haloes appear much more ubiquitous than they really are in images that are projected over thick slices.}. We also note that, while there is an abundance of vein-like structures in the DM maps for both $z = 0$ and $z = 2$, not all of them are filaments. Many are, instead, slices through walls that happen to intersect the projection plane at close to right angle. The two are difficult to distinguish in an individual projection plot, but we have tested this by scanning through the simulation box in a sequence of very thin slices ($\Delta = 0.5$ Mpc).

\begin{figure*}[t!]
  \includegraphics[width=\textwidth]{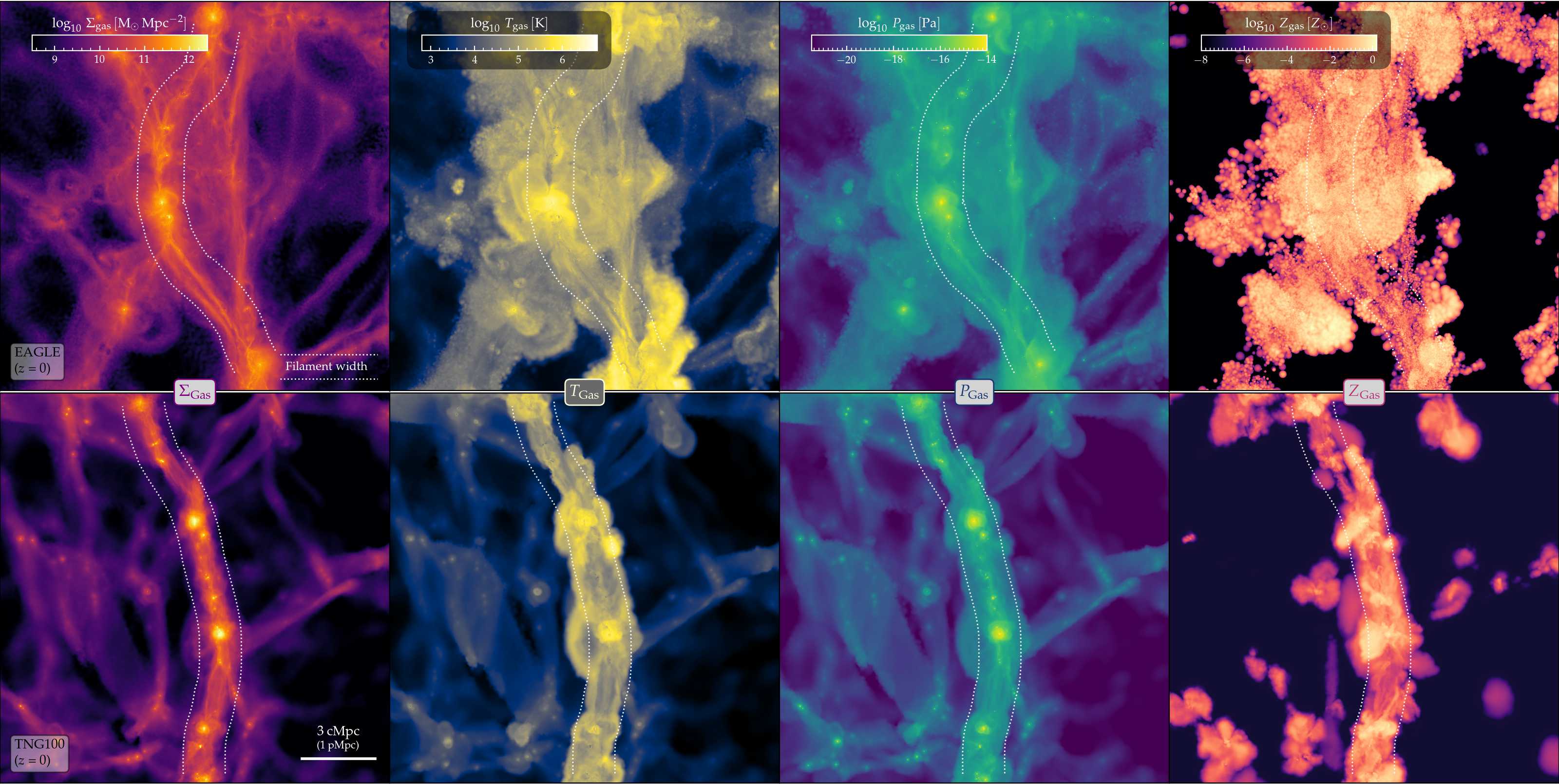}
  \caption{Detailed view of one representative filament at $z = 0$ in EAGLE (top) and TNG100 (bottom). From left to right, panels show maps of the gas density, temperature, pressure, and metallicity in a $16\times16\times1$ Mpc slice aligned with the best-fit plane of the filament. White dotted lines trace the edges of each filament as identified from the dark matter. There is abundant structure within the filaments, both along and perpendicular to the spine, with clear differences between the two simulations (see text for details).}
  \label{fig:17_single_filament_z0}
\end{figure*}

\begin{figure*}[t!]
  \includegraphics[width=\textwidth]{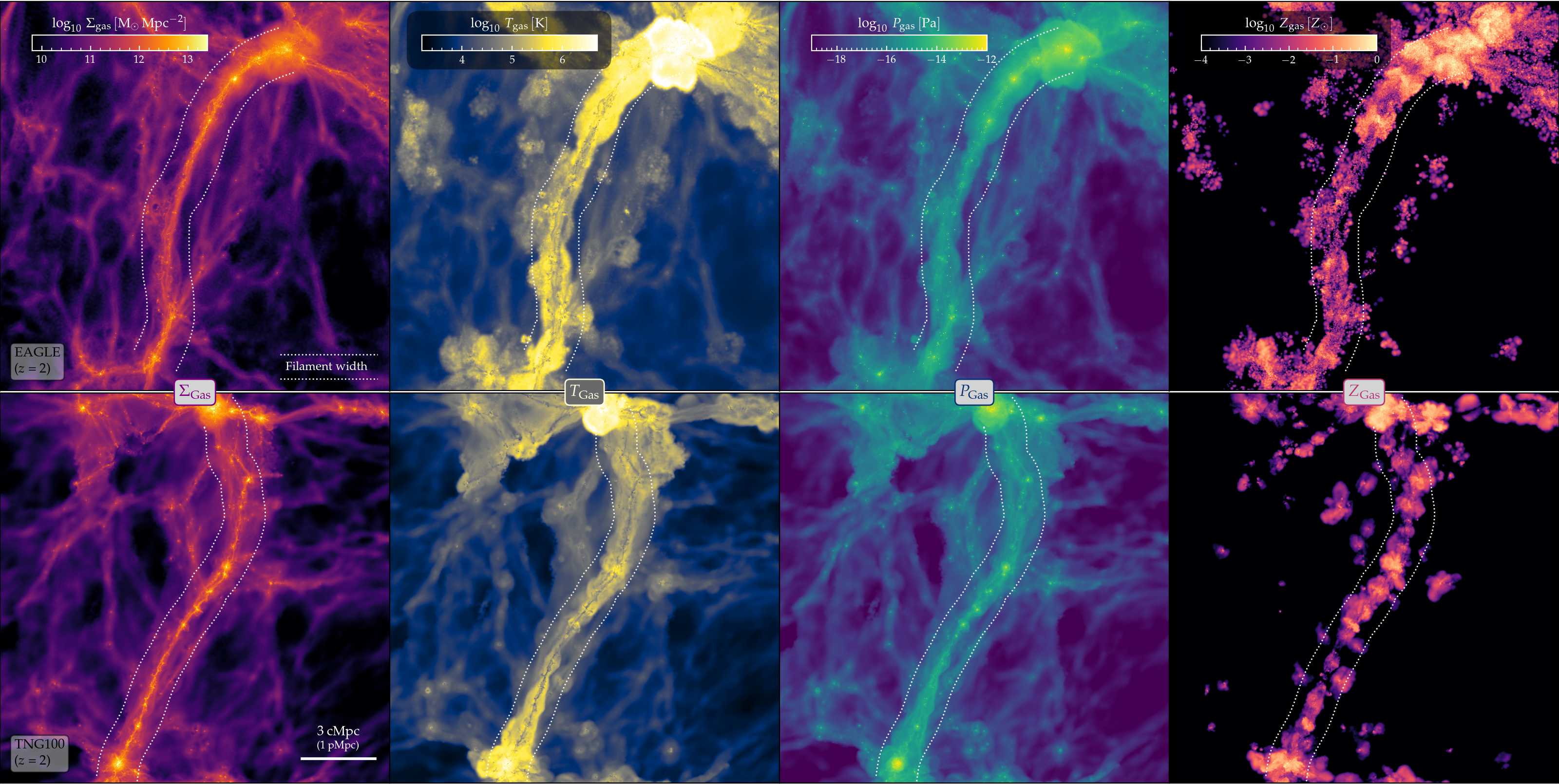}
  \caption{Same as Fig.~\ref{fig:17_single_filament_z0}, but showing two filaments at $z = 2$; the colour scales have been adjusted appropriately. Compared to $z = 0$, there is a more prominent concentration of pristine cool gas immediately around the spine in both simulations.}
  \label{fig:18_single_filament_z2}
\end{figure*}

Complementing this large-scale view, Figs.~\ref{fig:17_single_filament_z0} ($z = 0$) and \ref{fig:18_single_filament_z2} ($z = 2$) zoom in on one filament from each simulation. They show a smaller $16 \times 16 \times 1$ cMpc field, oriented along the best-fit plane to each filament. In order to be representative and inter-comparable, these four filaments were selected within narrow ranges in width (0.8--1.0 cMpc), length (15--17 cMpc) and overdensity within the central 300 ckpc (10--15); we also require their background overdensity (the `local floor' in the top panel of Fig.~\ref{fig:6_filament_density_profiles}) to be less than 0.75 to exclude filaments within large-scale overdensities.

\subsubsection{The gas structure at $z = 0$}
Focusing first on the large-scale view of gas density at $z = 0$ (Fig.~\ref{fig:15_cosmic_web_z0}, middle column), we find that it traces clear filament spines in both simulations, even for those filaments that appear highly clumpy in DM. This confirms our conclusion from the single EAGLE filament shown in Fig.~\ref{fig:1_filament_close_up} that gas is a clearer tracer of filament spines than DM.

In detail, there are significant qualitative difference between EAGLE and TNG100, despite their very similar DM structure. The TNG100 filaments are more clearly defined against their surroundings, and they are traced by many gas knots coinciding with small DM haloes that host $\mstar \lesssim 10^9\,\msun$ galaxies (not shown for clarity). In contrast, the EAGLE filaments appear smoother and more `blurred', indicating a more gradual decline of the gas density away from the spine (the average gas density profiles are, however, very similar between the two simulations, as shown in Fig.~\ref{fig:B1_filament_density_profiles_eagle_tng_z0}). The difference between smooth and knotted structure is particularly clear in the zoomed-in view of Fig.~\ref{fig:17_single_filament_z0} (left-hand column). EAGLE has the same abundance of small DM haloes along its filaments as TNG100, but these do not show up as prominently in the gas density map and must therefore be significantly less gas-rich (see also \citealt{Davies2020}). 

This different gas density structure is most likely the result of the very different sub-grid models for star formation feedback from low-mass galaxies embedded in the filaments. In EAGLE, newly formed stars cause strong stochastic explosions \citep{DallaVecchia_Schaye_2012,Schaye2015} that create powerful outflows from low-mass haloes \citep{Mitchell2020}. In contrast, star-forming gas in TNG100 drives (hydrodynamically decoupled) winds, whose launch speeds scale with the local DM velocity dispersion and are therefore lower in low-mass haloes \citep{Vogelsberger2013, Pillepich2018Model}. As shown by \citet{Nelson2019outflows}, this implementation leads to high gas outflow rates from low-mass galaxies only close to their centre, while the mass loading factor (mass of ejected gas per unit mass of stars formed) drops precipitously towards larger radii. This is in stark contrast to EAGLE---see the top-right panel of fig.~14 in \citet{Mitchell2020}---and explains both the lower gas content of low-mass haloes in EAGLE and its more blurred boundary of filament spines.

Compared to the view from gas density discussed above, the filaments have a rather different appearance in the temperature maps (right-hand panels of Fig.~\ref{fig:15_cosmic_web_z0} and second column of Fig.~\ref{fig:17_single_filament_z0}). In both simulations, the main filaments correspond to a broad band of hot gas, with a full width up to $\approx\,$5 Mpc. There is a sharp temperature contrast at the edge of the filament in both simulations, consistent with an accretion shock. Within the filaments, there is plenty of temperature substructure in the form of both cooler (especially near the spines) and also higher (in particular at larger radii) temperature `threads'. As these structures are similar in the two figures with different slice thickness, they are unlikely to be remaining projection effects. Prominent warm-hot bubbles exist around the locations of DM haloes, a clear signature of feedback from galaxies.

For the most massive haloes, for which feedback is dominated by AGN (e.g.~\citealt{Bower2017}), the bubbles are broadly consistent between both simulations. This is perhaps surprising, given that (efficient) AGN feedback in TNG100 is implemented as kinetic outflows, whereas EAGLE uses a thermal mode with a very high temperature increment ($\Delta T = 10^{8.5}$ K). The similar net effect on large scales indicates efficient conversion of either form of input energy: heated particles expand and drive outflows, whereas particles that have received a velocity kick will shock and therefore raise the surrounding temperature. Instead, the temperature map shows clear differences between EAGLE and TNG100 around smaller haloes, where stellar feedback dominates. The discrepancy in the temperature map is therefore consistent with our discussion above: stellar feedback in EAGLE drives stronger outflows from low-mass galaxies that heat up the gas in the filaments around them.

With this insight, we can now better understand the mass-weighted temperature profiles shown in Fig.~\ref{fig:14_temperature_profiles}. Their higher temperature compared to the length-weighted profiles is caused by the prominent hot haloes inflated by AGN feedback.  These haloes also lead to the rise in the outer profiles, because they are not just hot, but also have higher density than other regions at the same distance from the spine, so that they have a strong influence on the mass-weighted temperature. Meanwhile, the more pronounced central temperature dip in TNG100 results from the higher fraction of (cool-)gas rich haloes that are essentially absent in EAGLE but bias the average temperature near the cores of TNG100 filaments.

For the single-filament zoom-in views, we also show the gas pressure and metallicity in the right half of Fig.~\ref{fig:17_single_filament_z0} (and \ref{fig:18_single_filament_z2}). The filaments are significantly overpressurized, indicating that their gas is confined by the gravitational potential. Within the filaments, the pressure is fairly uniform (with the exception of the haloes, due to their even deeper potential wells), so that the different gas phases are in approximate pressure equilibrium.

The spatial distribution of the gas metallicity $Z_\mathrm{gas}$ is fundamentally different from the other three quantities. While there is a general increase of metallicity along the filament, this is all associated with outflows from embedded galaxies; $Z_\mathrm{gas}$ drops to almost zero between them, in both simulations. This indicates that there is very little gas mixing along the filament axis, unlike in haloes. The clear outflow patterns within filaments as traced by high-metallicity gas, coupled with the lack of strong gas mixing, also implies that galaxies are not passive `sinks' fed by gas streams along the filament axis, but play an active role in the gas dynamics of filaments. The main difference between EAGLE and TNG100 is the small-scale appearance of the metallicity maps (grainy vs.~smooth). This is due to the different hydrodynamics solvers: unlike \textsc{Arepo} gas cells, the SPH particles in EAGLE do not exchange metals between each other, so that there is a large scatter between the metal mass fraction of individual particles, and hence between neighbouring pixels.

\subsubsection{The gas structure at $z = 2$}
We now turn our attention to the $z = 2$ maps. Fig.~\ref{fig:16_cosmic_web_z2} depicts the same (co-moving) sub-volume of each simulation as Fig.~\ref{fig:15_cosmic_web_z0}, while Fig.~\ref{fig:18_single_filament_z2} shows two individual filaments that are not related to those in Fig.~\ref{fig:17_single_filament_z0}. Echoing what we noted in Sect.~\ref{sec:z2}, the DM filaments are noticeably thinner at $z = 2$ than at $z = 0$, and both haloes and filaments are more evenly spread out. Both are a direct consequence of the less progressed state of gravitational collapse at earlier cosmic times.

The gas density maps show a similar relation to the DM, and between the two simulations, as at $z = 0$: filaments are somewhat more clearly defined in gas than in DM; they stand out more sharply from the background in TNG100 than in EAGLE; and there is a greater abundance of small-scale gas-rich haloes in TNG100. One notable difference is that the gas density peaks more sharply near the spine, with a dense core that is much thinner than the DM filament width (dotted lines in Fig.~\ref{fig:18_single_filament_z2}).

The temperature maps are, however, very different from their $z = 0$ counterparts (right-hand column of Fig.~\ref{fig:16_cosmic_web_z2} and second column of Fig.~\ref{fig:18_single_filament_z2}). As expected from the significant offset between the temperature profiles of EAGLE and TNG100 (Fig.~\ref{fig:14_temperature_profiles}), the two simulations look strikingly different: EAGLE has an abundance of warm-hot gas with few regions dipping below $10^4$ K, whereas the TNG100 volume is dominated by cooler gas against which the filaments (and major haloes) stand out very clearly. An equally striking feature is the abundance of cool gas ($T \sim 10^4$ K) in the very centre of filaments---it stands out more noticeably for EAGLE in the large-scale view of Fig.~\ref{fig:16_cosmic_web_z2}, but is equally clear in the zoom-in maps of Fig.~\ref{fig:18_single_filament_z2}. Since we do not see cases where the cool gas is clearly offset from the spine, it must correspond to `cool cores' of gas in the centre of high-redshift filaments. Their extent corresponds closely to the aforementioned density cores, so that the structure is one of cool-dense gas surrounded by a warm-tenuous envelope.

Neither EAGLE nor TNG can model radiative gas cooling below $\sim$10$^4$ K (see e.g.~\citealt{Ploeckinger2020}), so that we cannot robustly predict the physical temperature of this gas, but it is clearly cooler than its surroundings. As seen from the pressure maps (third column of Fig.~\ref{fig:18_single_filament_z2}), both phases are approximately in pressure equilibrium and overpressurized compared to the surrounding cooler and less dense background. This agrees with the recent work of \citet{Lu2024}, who analysed filaments in a higher-resolution simulation with the TNG model at $z \sim 4$ and also found an inner `stream' zone dominated by cool ($T \sim 3 \cdot 10^4$ K) gas from an isobaric cooling flow. Our results imply that this structure is also formed at the lower resolution accessible to current simulations of representative volumes, and that it is not unique to the TNG model.

Similar to $z = 0$, metallicity is only enhanced around galaxies, while much of the gas around the spine is at $\lesssim$10$^{-2}\,Z_\odot$ (right-most column of Fig.~\ref{fig:18_single_filament_z2}). 
The cool gas discussed above is therefore nearly primordial, so that the most plausible reason for its efficient cooling is not metal enrichment but its significantly higher physical gas density compared to $z = 0$: as we had seen in Fig.~\ref{fig:13_filament_properties_z2}, the overdensity in the centre of filaments is similar at $z = 0$ and $z = 2$, but as the Universe was only one third of its present-day size, the background density---and hence also the physical density of gas around the filament spines---was a factor $3^3 = 27$ higher. The gas cooling rate depends on the square of the density, and was therefore orders of magnitude higher at $z = 2$. In other words, gas in high redshift filaments could cool efficiently, and from there directly accrete onto galaxies \citep{Dekel2006}. This is the origin of the temperature downturn in the central filament regions seen in Fig.~\ref{fig:14_temperature_profiles}.

\section{Identifying filaments from galaxies instead of DM}
\label{sec:galaxy-filaments}

Our analysis above is based on filaments identified from the DM density. As discussed in the Introduction and in Sect.~\ref{sec:filaments:multiscale}, this approach is preferable to a galaxy-based identification on physical grounds, but it is currently impossible to apply to observations. We now compare the results of our DM-based method to the more observational approach of applying DisPerSE to the simulated galaxy catalogue. This not only serves to relate our findings better to observational works, but also to the large body of previous simulation studies that have identified filaments based on galaxies. Broadly similar comparisons have been performed by \citet{Laigle2018} and \citet{Zakharova2023}, which we discuss in Sec.~\ref{sec:galfil:comparison}. For simplicity, we only use EAGLE here, and only look at $z = 0$, but would expect very similar results for TNG100 given the close agreement between the DM-filaments of the two simulations.

There are considerably fewer choices to make for the galaxy-based filament finding than in our fiducial DM-based method. One is the selection of input galaxies. Since our approach above aimed at identifying filaments that host galaxies with $\mstar > 10^9\,\msun$, we select our tracer galaxies here above the same threshold. Observationally, this cut-off mass is at the lower end of what is currently accessible to systematic high-completeness spectroscopic surveys beyond the very local Universe\footnote{Lower mass limits can be achieved for star-forming galaxies, but the filaments may then be biased so that the interpretation relies heavily on the (limited) accuracy with which current simulations can predict whether or not a galaxy is quenched.}---GAMA, for instance, is complete down to $\mstar = 10^9\,\msun$ out to $z = 0.06$ \citep{Davies2016}. The second parameter is the DisPerSE persistence threshold, which can in this case be expressed in terms of the statistical significance of the critical points at which filaments end. To make the result as comparable as possible to our DM-based method, we select a rather liberal threshold of $3\sigma$. As appropriate for our setup, we use periodic boundary conditions. The resulting filaments are smoothed once to remove artificial small-scale wiggles, but this has barely any effect here since the vast majority of filaments are traced by $\leq 5$ sampling points.

\subsection{Visual comparison}

\begin{figure}
  \includegraphics[width=\columnwidth]{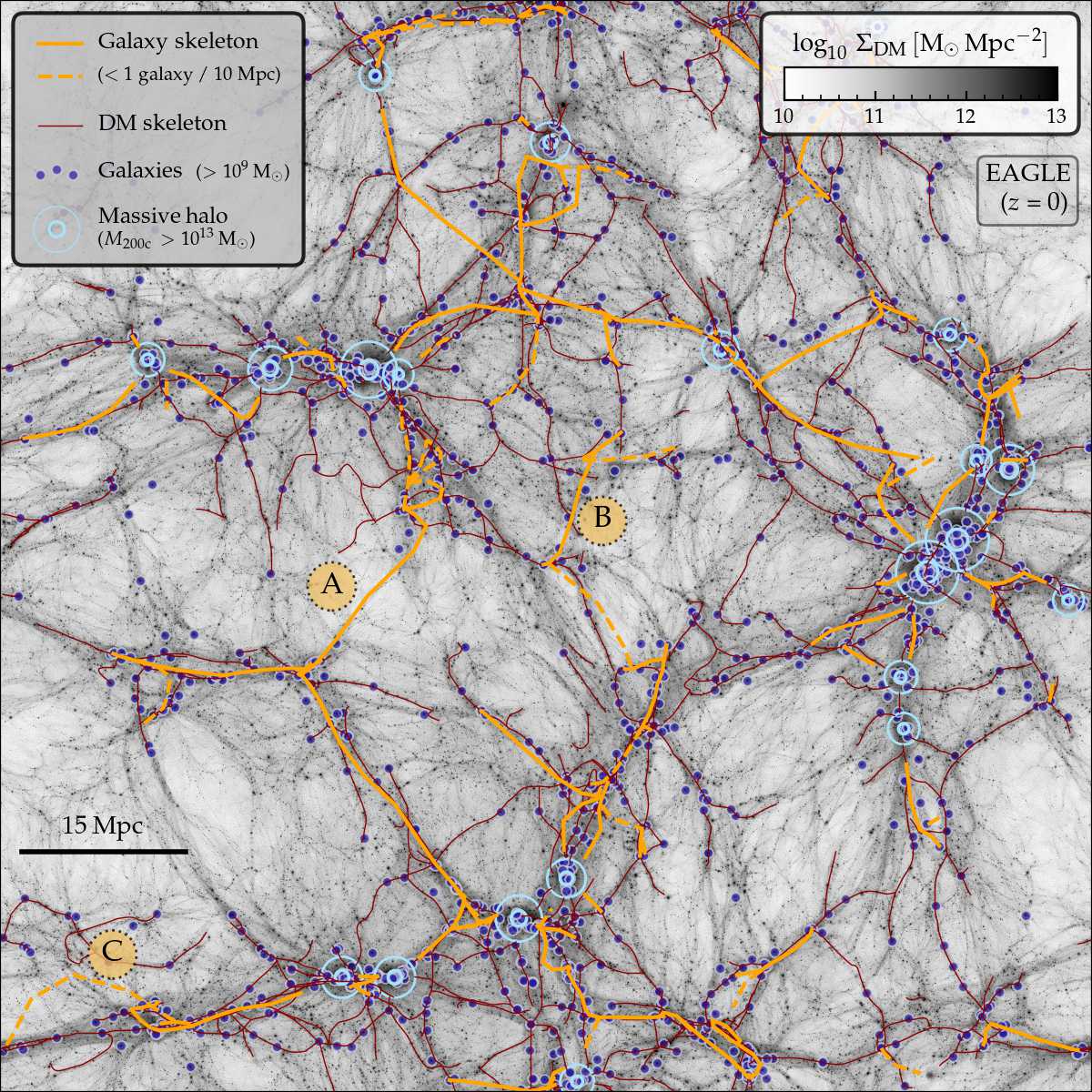}
  \caption{Filament network identified from galaxies (orange) compared with our fiducial reconstruction from the DM density field (dark red), for EAGLE at $z = 0$, showing the same $100\times100\times15$ Mpc slice as in Fig.~\ref{fig:2_overview}. The greyscale background represents the DM surface density, while indigo points and light blue circles denote galaxies with $\mstar > 10^9\,\msun$ and haloes with $M_\mathrm{200c} > 10^{13}\,\msun$, respectively. The galaxy-based network broadly captures the thicker DM-filaments, but misses most finer structures and suffers from a number of artefacts, examples of which are labelled as `A', `B', and `C' (see text for details).}
  \label{fig:19_galaxy_skeleton_image}
\end{figure}

A visual comparison of the DM- and galaxy-identified filament networks is shown in Fig.~\ref{fig:19_galaxy_skeleton_image}. The setup is equivalent to the left-hand panel of Fig.~\ref{fig:2_overview}: a 15 Mpc slice of the full simulation box (100 Mpc side length) with the background showing the DM density, light blue circles the location of haloes with $M_\mathrm{200c} \geq 10^{13}\,\msun$, and small indigo dots the coordinates of galaxies with $\mstar > 10^9\,\msun$. The galaxy-based filaments are drawn with thick orange lines, the DM-based ones with thin dark red ones. We do not differentiate between filaments of different thickness here, because the sparseness of galaxies makes it impossible to construct a meaningful galaxy density profile for individual filaments (those shown in Fig.~\ref{fig:9_filament_density_profiles} were based on stacks of 600 filaments each).

Both galaxy- and DM-filament networks cover similar parts of the map, and mostly avoid the prominent void regions. In detail, however, there are clear differences between the two. The galaxy-filament network (orange) is less extended overall: its total length amounts to only 28 per cent of the DM-filaments (6\,244 vs.~22\,297 Mpc in the full simulation volume). Many of the finer and thinner filaments that are identified from DM (dark red) have no galaxy-based counterpart, including ones that host galaxies above our $10^9\,\msun$ stellar mass threshold. Moreover, the galaxy filament network has some noteworthy artefacts, examples of which are labelled with letters in orange circles in Fig.~\ref{fig:19_galaxy_skeleton_image}: bridges across empty regions of space (A); `pseudo-filaments' composed of galaxies that actually belong to multiple different DM-filaments (B); and spines that are well offset from their DM counterparts (C). Some of the more glaring problems can be ameliorated by excluding galaxy-filaments with a very low galaxy density along their spine (dashed lines; fewer than 1 galaxy per 10 Mpc within 200 kpc around the spine), but even with such a cut many of the aforementioned artefacts remain.

While these shortcomings might, in principle, be intrinsic to DisPerSE (notwithstanding its good result when applied to DM), they can all be directly traced to the very sparse nature of galaxies as cosmic web tracers, as discussed in Sect.~\ref{sec:filaments:multiscale}. We have verified that lowering the persistence threshold to $2\sigma$ mostly increases the number of spurious filament identifications, while many of the fainter DM filaments are still unmatched. Lowering the galaxy mass threshold to $10^8\,\msun$ helps with the latter, but also adds many more spurious filaments so that the network remains substantially different from the DM-based one. For the sake of conciseness, we do not show figures from these additional tests.

\subsection{Galaxies around DM- and galaxy-filaments}

As a more quantitative comparison of the galaxy- and DM-filaments, we test how the distance of galaxies from their nearest galaxy-filament ($d_\mathrm{gal}$) relates to their true environment, which we parameterise as (i) the distance to a DM-filament ($d_\mathrm{DM}$) and (ii) the width of the DM-filament (if any) to which a galaxy belongs. We use these metrics, rather than the distance between the filament skeletons \citep{Laigle2018,Zakharova2023}, on account of their unambiguous definition and direct relation to observational studies of filament galaxies, for which $d_\mathrm{gal}$ is a key measurable parameter. Specifically, we measure $d_\mathrm{gal}$ and $d_\mathrm{DM}$ as the minimum perpendicular distance to all segments in the respective filament network. We define a galaxy as `belonging' to a DM-filament if the minimum distance between the two is less than the filament width. If more than one filament satisfies this criterion for a given galaxy, we choose the one with the smallest separation; if there are none, we classify the galaxy as `not in an identified filament'. Rather than labelling these galaxies as `field', `wall' or `void', this terminology makes it explicit that they are (typically) still part of a filamentary structure, albeit one that is too tenuous to make our identification cut.

\begin{figure*}
  \includegraphics[width=\textwidth]{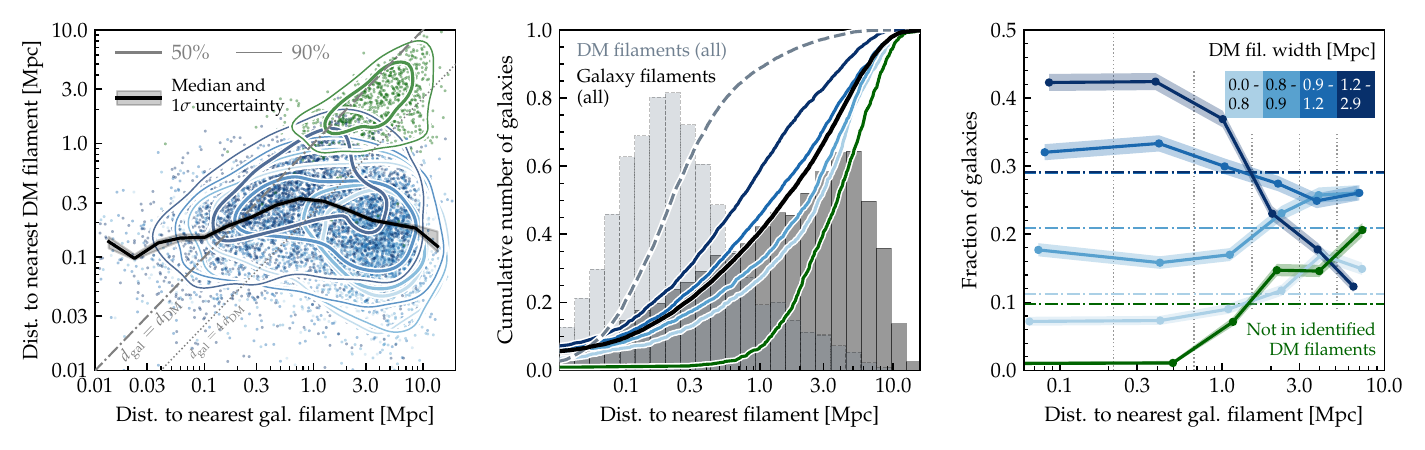}
  \caption{Association of galaxies ($\mstar > 10^9\,\msun$, excluding those near a group or cluster) to filaments identified from galaxies and from dark matter (DM). Shades of blue indicate DM-filaments of different width (see top-left corner of the left-hand panel), green denotes galaxies that do not belong to any identified DM filament. \emph{Left:} Comparison of distances to the spine of the nearest DM and galaxy filament ($d_\mathrm{DM}$, $d_\mathrm{gal}$). Points show individual galaxies, with correspondingly coloured lines enclosing the central 50 and 90 per cent in each environment. The thick black line and shaded band traces the running median of $d_\mathrm{DM}$ in bins of $d_\mathrm{gal}$ and its 1$\sigma$ uncertainty. \emph{Middle:} Cumulative distribution of galaxy distances from their nearest galaxy-filament spine. The black line represents the full galaxy sample, corresponding to the dark grey histogram in the background. Coloured lines split the galaxy population into the same five environments as in the left-hand panel (histograms omitted for clarity). For comparison, the distribution of galaxy distances from their nearest DM-filament is shown by the grey line and histogram. \emph{Right:} Fraction of galaxies in each of the five environments as a function of distance from their nearest galaxy-filament spine. Shaded bands indicate binomial 68 per cent confidence intervals \citep{Cameron2011}. Thin dotted vertical lines delineate the five distance bins; each point is plotted at the median distance of the galaxies that it represents. Horizontal dash-dotted lines show the overall fraction of galaxies in each environment. Distances to the two filament networks are only weakly correlated, but galaxies closer to a galaxy-filament are generally more likely to be part of a thick DM-filament.}
  \label{fig:20_galaxy_distance_comparison}
\end{figure*}

The left-hand panel of Fig.~\ref{fig:20_galaxy_distance_comparison} compares the distances of galaxies ($M_\mathrm{star} > 10^9\,\msun$) to their nearest DM- and galaxy-filament. In analogy to our filament masking step (Sect.~\ref{ssec:masking}), we exclude any galaxies that lie within 3 $r_\mathrm{200c}$ of a $M_\mathrm{200c} > 10^{13}\,\msun$ halo, because these are more appropriately classified as group or cluster galaxies. Galaxies that belong to a filament are grouped according to the width quartiles of that filament, represented by different shades of blue (see the legend in the right-hand panel); those not in any identified (DM) filament are shown in green. The running median of $d_\mathrm{DM}$ in bins of $d_\mathrm{gal}$ is traced by the black line as a quantitative measure of the correlation between the two distances.

While $d_\mathrm{gal}$ and $d_\mathrm{DM}$ span broadly the same range ($\sim$10 kpc--10 Mpc, but see below), they are far from equivalent. Half the galaxies are loosely clustered along the equality line ($d_\mathrm{gal} < 4\,d_\mathrm{DM}$, with a standard deviation of 0.33 dex in $d_\mathrm{gal}/d_\mathrm{DM}$), while the other half form a clump around $d_\mathrm{gal} \approx 3\,\mathrm{Mpc}$ and $d_\mathrm{DM} \approx 0.2\,\mathrm{Mpc}$. Different DM environments overlap substantially, but in tendency galaxies in the thickest DM-filaments, and those outside any DM-filament, are most common in the first set while thin DM-filaments dominate the clump of outliers towards the bottom-right corner. The latter is therefore due to filaments that are missed in the galaxy-based identification, which is unsurprisingly more common for thinner ones. In combination, this two-component structure results in a modest correlation between $d_\mathrm{gal}$ and $d_\mathrm{DM}$ for $0.03 < d_\mathrm{gal} < 0.7\,\mathrm{Mpc}$ (Spearman rank-order correlation coefficient $r = 0.34$, standard deviation of 0.43 dex), but the trend then inverts for larger $d_\mathrm{gal}$ with a mild anti-correlation and even larger scatter. Put simply, $d_\mathrm{gal}$ is a moderately good proxy for $d_\mathrm{DM}$ near a galaxy-filament spine ($d_\mathrm{gal} \lesssim 0.7\,\mathrm{Mpc}$) but not at larger values.

The distributions of $d_\mathrm{gal}$ for the five environments can be compared more clearly in the middle panel. Although all are quite broad (with the exception of galaxies not in identified DM filaments), different environments are clearly and systematically offset from each other, with a median of 3.8 Mpc for galaxies not in any identified DM filaments (green), 2.7 Mpc for galaxies in the least dense ones (light blue), and only 0.7 Mpc for galaxies in the densest DM filaments (dark blue). Even for this last category, $d_\mathrm{gal}$ is still systematically larger than $d_\mathrm{DM}$ for the full galaxy sample, which is concentrated around a well-defined peak at only 0.2 Mpc; galaxies are typically closer to a DM- than a galaxy filament. A few per cent of galaxies are outliers from these distributions and lie (within floating point accuracy) exactly on the spine. Most likely, this is because DisPerSE uses the positions of galaxies to define the cell structure on which filaments are identified.

In the right-hand panel of Fig.~\ref{fig:20_galaxy_distance_comparison}, we compare directly the mix of true environments at different distances from a galaxy-filament. At the smallest $d_\mathrm{gal}$ ($<\,0.2\,\mathrm{Mpc}$) there is a clear separation between the five environments, with only 1 per cent of galaxies not in identified filaments but 42 per cent in the thickest quartile. These fractions are almost the same in the next $d_\mathrm{gal}$ bin, since we are still probing distances smaller than the width of most filaments. The different fractions then converge steadily towards larger radii, with all five environments accounting for between 11 and 26 per cent of galaxies in the outermost bin ($d_\mathrm{gal} > 4$ Mpc). The radial trend in the fractions, and the excess (deficit) of high- (low-)density environments in the innermost bins compared to the average, confirms that $d_\mathrm{gal}$ can be used statistically as an environment proxy. We stress, however, that there is substantial overlap between DM-filament widths in all $d_\mathrm{gal}$ bins, and that even at $d_\mathrm{gal} < 200$ kpc less than half of all galaxies belong to the thickest quartile.

\subsection{Density profiles}

\begin{figure}
  \includegraphics[width=\columnwidth]{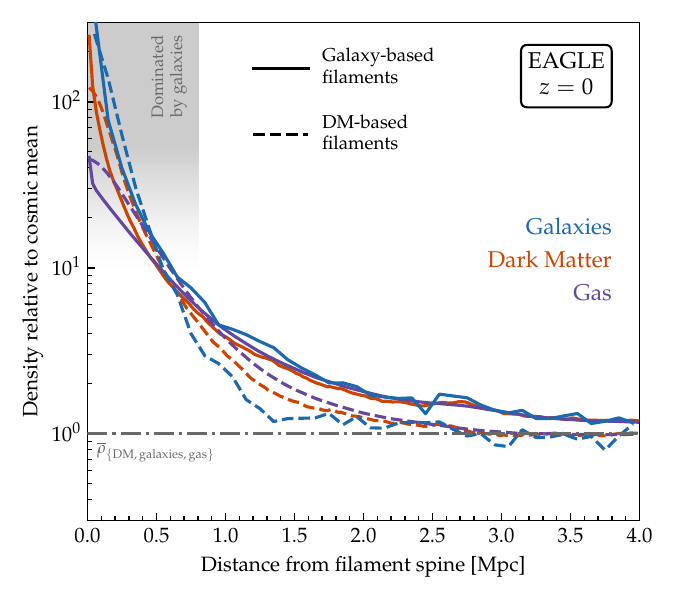}
  \caption{Density profiles of galaxies (blue), dark matter (orange) and gas (purple) around filaments reconstructed from the galaxy distribution (solid lines) of EAGLE at $z = 0$. Equivalent profiles around a matched sample of dark matter based filaments are shown as dashed lines for comparison. For all three components, the profiles around galaxy-based filaments agree quite well with their DM counterparts in the central $\approx\,$0.5 Mpc, but are significantly more extended at larger radii.}
  \label{fig:21_density_profiles_galaxyfilaments}
\end{figure}

As a final test of the galaxy-based filament network, we compare its density profiles to those around DM-filaments in Fig.~\ref{fig:21_density_profiles_galaxyfilaments}. Similar to Fig.~\ref{fig:9_filament_density_profiles}, we construct overdensity profiles for galaxies, dark matter, and gas. As in Fig.~\ref{fig:19_galaxy_skeleton_image}, we cannot differentiate between galaxy-filaments of different width, but a comparison between all galaxy- and DM-filaments would be biased because many of the thinner filaments are missed in the former. We therefore select the $N=804$ thickest DM-filaments (width $\geq 1.095$ Mpc), such that their total length matches that of the (full) galaxy filament network for an approximately fair comparison.

In the central $\sim$500 kpc, the profiles for all three components (galaxies, DM, gas) around galaxy filaments agree quite closely with their DM-filament counterparts, to better than the offset between the three components. However, we have already seen (Fig.~\ref{fig:9_filament_density_profiles}) that this part of the profiles is dominated by galaxies and their satellites; the agreement is therefore not too surprising. At larger radii, however, the galaxy-filament profiles drop more slowly, and in contrast to the DM-filament profiles remain well above the cosmic mean beyond 3 Mpc. The less precise placement of filament spines with galaxies as tracers therefore leads to the illusion of a diffuse extended structure\footnote{It is worth noting that this is not due to residual bias of galaxy-filaments to thicker structures---which simply means that we do not probe around less prominent filaments---and also not due to the inclusion of galaxies or matter in different physical filaments, which occurs also in the DM-based profiles.}. Most likely, this explains why we find much thinner widths for filaments than e.g.~\citet{Galarraga-Espinosa2020} and a different redshift evolution of filament widths than \citet{Wang2024}.

\subsection{Comparison to other studies}
\label{sec:galfil:comparison}

Ours is not the first comparison between galaxy- and DM-based filaments, although to our knowledge no previous works have compared the two in as much detail for a representative part of the cosmic web. \citet{Laigle2018} identify filaments via DM and galaxies in the Horizon-AGN simulation using DisPerSE and compare the projected distance between the two (see their appendix C and fig.~C1). They find a median (projected) offset of 0.3--0.8 Mpc between the two, depending on the galaxy mass limit and direction of matching (DM- to nearest galaxy- vs.~galaxy- to nearest DM-filament). This is broadly consistent with our Fig.~\ref{fig:19_galaxy_skeleton_image}, keeping in mind that we likely include thinner filaments in our DM catalogue than \citet{Laigle2018}. Nevertheless, our comparison makes it clear that even filament networks that overlap quite closely along much of their length can have very different shapes, profiles, and associations with galaxies.

\citet{Zakharova2023} performed a detailed study of DM- vs.~galaxy-filaments around massive galaxy clusters, also using DisPerSE. Different from our approach (and from \citealt{Laigle2018}), they calibrate the persistence threshold for the DM filament network such that its total length is the same as for the galaxy-based one. While this should minimize the difference between the two approaches, they find---like us---only limited agreement between the DM- and galaxy filaments (see their figs.~1 and 6), with almost half of their DM-filaments having no clear galaxy-based analogue. Like \citet{Laigle2018}, \citet{Zakharova2023} only compute distances between pairs of DM- and galaxy-filaments---again finding a rather broad range of $\sim$0.1--10 Mpc---rather than between filaments and galaxies as we did. Nevertheless, the general agreement between our results and theirs suggests that the discrepancy between DM- and galaxy-based filaments is generic, affecting both the representative filament population that we study and the extreme subset around massive clusters analysed by \citet{Zakharova2023}.

We stress, however, that our comparison has been far from exhaustive. It is conceivable that more sophisticated filament finders and/or density estimators may be able to extract a more accurate filament network from galaxies. For example, \citet{Bonnaire2020} show that their (graph-based) `T-Rex' filament finder avoids some of the artefacts seen in the DisPerSE galaxy-filament network. Even within the DisPerSE framework, \citet{Hasan2024} find that a different way of calculating the density field from the galaxy coordinates leads to closer agreement with the underlying DM density, and hence to a more realistic filament network, although they do not compare to a DM-based version like we have done. Our results indicate that developments of this kind can improve our observational view of filaments and their galaxies, although more work needs to be done to fully assess their capabilities and limitations.

\section{Summary and conclusions}
\label{sec:summary}

Filaments harbour the majority of galaxies in the Universe, but their non-trivial identification and complex multi-scale structure limits our understanding of what they are and how they influence the evolution of galaxies. Using the publicly available simulations EAGLE Ref-L0100N1504 (`EAGLE') and IllustrisTNG100 (`TNG100'), we have built filament catalogues based directly on the simulated dark matter (DM) density field at redshifts $z = 0$ and $z = 2$, using the DisPerSE algorithm with careful discussion of all involved parameters. We use DM density profiles around filaments to compute a physically-motivated measure of their radial widths, analyse their structure using a combination of parametric and image-based diagnostics, and compare to the more observation-compatible approach of using galaxies as tracers. The main results from this work can be summarized as follows:

\begin{enumerate}
    \item With appropriate density smoothing and post-processing, DisPerSE can identify a network of filaments that closely resembles the relevant features of the underlying DM density field (Fig.~\ref{fig:2_overview}). The result depends sensitively on the input parameters, in particular the DM smoothing length and persistence threshold, and to a lesser extent on the smoothing of the extracted filaments. All of these must be calibrated for a given scientific objective, in our case to capture filaments down to the scale where they host most $\mstar = 10^9\,\msun$ galaxies (Fig.~\ref{fig:4_disperse_variations}).
    
    \item Along most of their length (i.e.~in a median-averaged sense), filaments reach typical DM overdensities of only $\sim$10 in their centre. On their outskirts, DM densities drop well below the cosmic mean, but because most of the Universe is highly underdense, even these modest underdensities represent local DM enhancements (Fig.~\ref{fig:6_filament_density_profiles}).
    
    \item Filaments are a heterogeneous class of environments. At $z = 0$, those that we identify span a factor $\approx\,$7\,$\times$ in width (0.54--2.9 Mpc), and factors $\approx\,$100$\times$ in central DM overdensity ($\approx\,$1--100) and length ($\approx\,$0.5--50 Mpc). The distributions of these parameters are near-identical between EAGLE and TNG100 (Fig.~\ref{fig:7_filament_properties}), but they are only weakly correlated between each other (Fig.~\ref{fig:8_filament_property_correlations}). This diversity limits the meaning of `filament' without further distinction or details as an environment label.
    
    \item The \emph{mean} overdensity profiles of galaxies, gas, and DM around filaments agree closely, with galaxies slightly more and gas slightly less concentrated around filament spines than DM. Their central parts are biased high by the presence of highly overdense substructures and reach overdensities of $\gtrsim 100$. All average profiles converge to the cosmic mean less than 3 Mpc from the spine, indicating that filaments are relatively thin structures. (Fig.~\ref{fig:9_filament_density_profiles}).
    
    \item The cores of most filaments are strongly substructure dominated; only 21 per cent (median) of DM mass is in a diffuse background within 100 kpc from the spine. Towards the outskirts, the diffuse fraction rises to $\approx\,$75 per cent. This is the opposite of groups or clusters, where the diffuse fraction increases to $\gtrsim\,$90 per cent towards the centre (Fig.~\ref{fig:10_filament_smoothness}). This clumpy, multi-scale nature of filaments complicates their definition, extraction from simulations, and comparison across different methodologies.
    
    \item The cosmic web evolves strongly with redshift. At $z = 2$, density contrasts are lower, particularly due to less developed underdensities in voids, while filaments are less prominent and less strongly aligned towards nodes than at $z = 0$. At even higher redshift, DM forms very thin `proto-filaments' with clear visual differences from their $z = 0$ descendants. (Fig.~\ref{fig:11_cosmic_web_evolution}). This evolution must be reflected in the parameters used to extract filaments at $z \gg 0$ if the result is to remain physically meaningful (Fig.~\ref{fig:12_skeleton_z2}).
    
    \item Compared to $z = 0$, filaments at $z = 2$ have almost the same distribution in (co-moving) lengths, very similar overdensities, but are $\approx\,$30 per cent thinner even in co-moving units (Fig.~\ref{fig:13_filament_properties_z2}). This demonstrates that filaments remain coupled to cosmic expansion and is consistent with inside-out growth due to the accretion of mass from their surroundings.
    
    \item Gas in filaments is hotter than around them. At $z = 0$, the length-weighted average temperature near the spine is $\sim$10$^5$ K, and $\sim$3$\cdot 10^5$ K in a mass-weighted sense due to the influence of hot and overdense substructures. Temperatures are higher for thicker filaments, agree broadly between EAGLE and TNG100, and only asymptote to a $\approx\,3\cdot 10^4\,\mathrm{K}$ floor at $\gtrsim\,$4 Mpc for the thickest filaments. At $z = 2$, the temperature profiles drop in the central $\lesssim\,$500 kpc, differences between filament widths are much smaller than at $z = 0$, but temperatures predicted by EAGLE are systematically $\approx\,$2$\times$ higher than for TNG100 (Fig.~\ref{fig:14_temperature_profiles}).
    
    \item Despite close agreement in DM, there are significant differences in the gas structure of filaments between EAGLE and TNG100. In EAGLE, the gas density is smooth within filaments and blurs into the background, while TNG100 filaments are sharply offset from their surroundings and contain a myriad of gas-rich low-mass haloes. Both simulations predict relatively sharp temperature edges around thick filaments and prominent hot bubbles around massive haloes. In contrast, thin filaments are inflated with warm ($\sim$$3\cdot 10^4$--$10^5$ K) gas in EAGLE while their temperatures only reach $\lesssim\,$10$^4$ K, and only in a narrow region around their spines, in TNG100 (Figs.~\ref{fig:15_cosmic_web_z0} and \ref{fig:16_cosmic_web_z2}). These differences are plausibly related to the different stellar feedback implementations of the two simulations, despite their agreement on galaxy properties.
    
    \item Detailed inspection of individual filaments confirms that the DM-defined filament width closely corresponds to edges in their gas structure, and highlights plentiful substructure within filaments. The temperature is non-monotonic even at $z = 0$, with the highest temperature typically near the edge, consistent with an accretion shock. The metallicity distribution is highly inhomogeneous, ruling out significant gas mixing along the filament axis. At $z = 2$, gas forms a thin, highly overdense, and cool core near the spine (Figs.~\ref{fig:17_single_filament_z0} and \ref{fig:18_single_filament_z2}), plausibly formed through efficient in-situ cooling due to the much higher physical densities than at $z = 0$.
    
    \item Using galaxies as tracers only allows identifying the most prominent filaments, with a variety of artefacts; the result is not a close analogue of our default DM-based approach (Fig.~\ref{fig:19_galaxy_skeleton_image}). The distance of a galaxy to its nearest galaxy-filament is only weakly correlated with distance from a DM-filament, but it can serve as a statistical proxy of the galaxy's DM-filament width and hence its local environment (Fig.~\ref{fig:20_galaxy_distance_comparison}). As a result of spine miscentering, density profiles around galaxy-filaments decay more slowly and are still well above the cosmic mean at 4 Mpc, giving the illusion that filaments are much thicker than they really are (Fig.~\ref{fig:21_density_profiles_galaxyfilaments}).  
    
\end{enumerate}

At a high level, our analysis demonstrates that filaments (and the cosmic web more generally) are complex structures for which simple models based on cylindrical symmetry only provide a limited description: filaments are not simply stretched-out versions of galaxy clusters. We have demonstrated that taking these complexities into account can lead to a rich and detailed view of filaments and the galaxies embedded in them, with results that often contradict those derived with other approaches. The very close agreement between EAGLE and TNG100, two independently developed and calibrated simulations, on the structural parameters of filaments strongly suggests that these are robust predictions that can be applied to observations, at least under the fundamental assumption of a $\Lambda$CDM cosmology. The same is not true for baryon properties, however, where discrepancies between the two simulations warrant careful validation across multiple simulations and/or observations, and caution against over-interpreting their predictions.

With the filament catalogues that we have built in this work, and the insight into their DM and gas structure, we are now able to investigate the co-evolution between filaments and their galaxies. This will be the topic of forthcoming works in this series: one in which we analyse the predicted properties of galaxies as a function of their filament environment, and one focused on the assembly of filaments and their galaxies over cosmic time. Together these will shed important new light on how the properties of filaments and their galaxies are driven by their mutual co-evolution.

\begin{acknowledgements}
The authors gratefully acknowledge helpful discussions with Alfonso Arag\'{o}n-Salamanca, C\'{e}line Gouin, Meghan Gray, Katarina Kraljic, Ulrike Kuchner, Clotilde Laigle, Nicolas Malavasi, Christophe Pichon, and Linda Tacconi. YMB acknowledges support from UK Research and Innovation through a Future Leaders Fellowship (grant agreement MR/X035166/1). This work was supported by the Swiss National Science Foundation (SNSF) under funding reference 200021\_213076. This work used the DiRAC@Durham facility managed by the Institute for Computational Cosmology on behalf of the STFC DiRAC HPC Facility (www.dirac.ac.uk). The equipment was funded by BEIS capital funding via STFC capital grants ST/K00042X/1, ST/P002293/1, ST/R002371/1 and ST/S002502/1, Durham University and STFC operations grant ST/R000832/1. DiRAC is part of the National e-Infrastructure. This research has made use of NASA's Astrophysics Data System. All figures in this paper were produced using the \textsc{Astropy} \citep{astropy:2022} and \textsc{Matplotlib} \citep{Hunter2007} Python packages.
\end{acknowledgements}

%
%

\bibliographystyle{aa}
\bibliography{filaments}

\begin{appendix}
\section{Comparison of smoothing styles}
\label{app:logsmoothing}
\begin{figure}
  \includegraphics[width=\columnwidth]{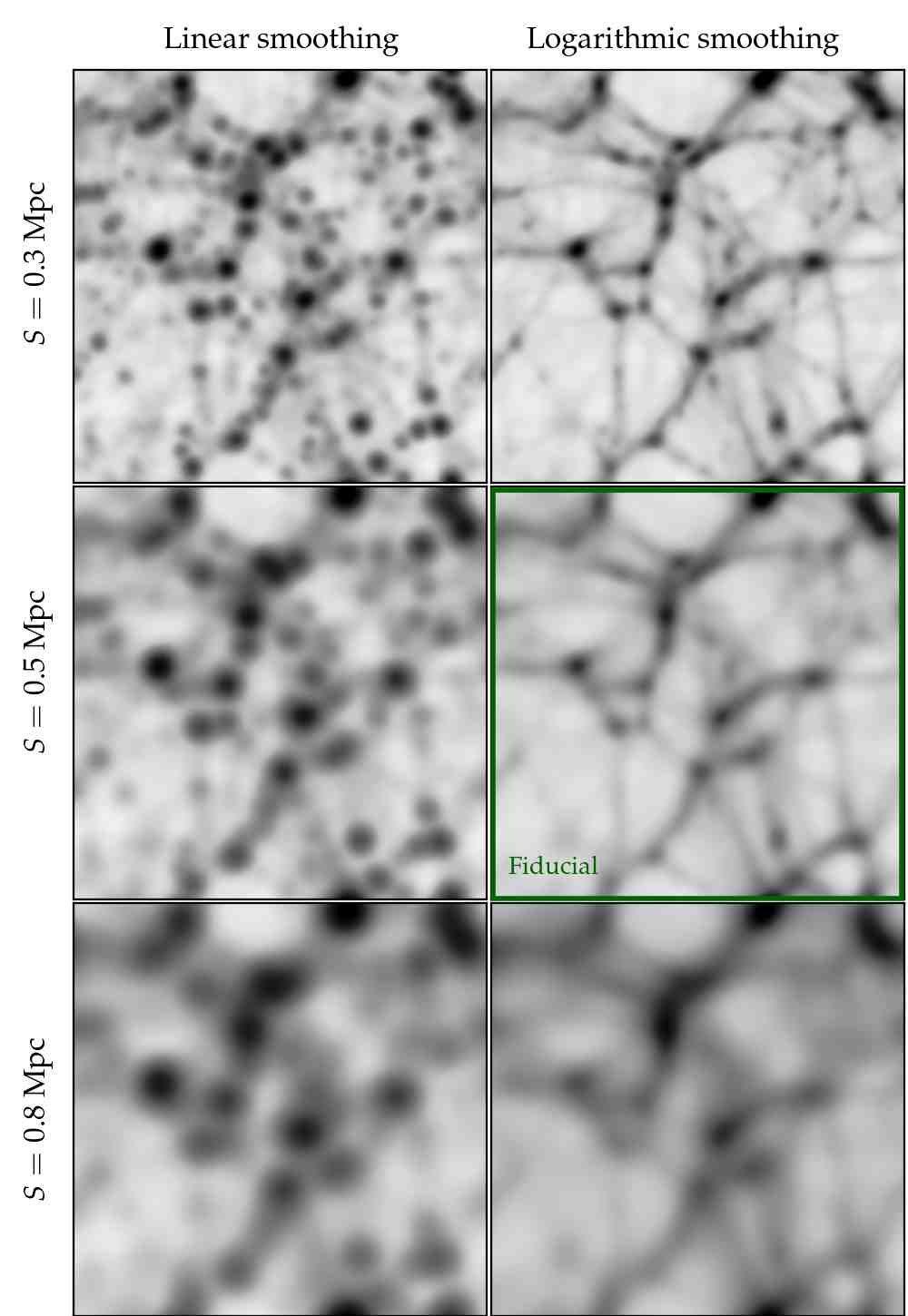}
  \caption{Comparison of the projected DM density field in a (25 cMpc)$^3$ volume smoothed in linear (left column) and logarithmic space (right column), at three identical smoothing lengths $S$. While linear smoothing mainly blurs, and hence emphasizes, the nodes, logarithmic smoothing brings out the fainter features such as filaments. The middle-right panel, framed in green, corresponds to our fiducial choice.}
  \label{fig:A1_smoothing_type_comparion}
\end{figure}

Fig.~\ref{fig:A1_smoothing_type_comparion} demonstrates the difference between smoothing the DM density field in linear and logarithmic space. Each of the six panels shows the projected DM density in the same (25 cMpc)$^3$ volume of the EAGLE-Ref25 simulation but smoothed in different ways. In the three left-hand panels, the linear density field is smoothed, whereas we smooth its logarithm for the three right-hand panels. Each row uses the same smoothing length, as indicated on the left edge. There is a clear qualitative difference between the two methods, with linear smoothing mainly blurring out---and hence emphasizing---the nodes, whereas logarithmic smoothing emphasizes the fainter filamentary features. This corresponds to a better ability of DisPerSE to recover the filaments (not shown here), which is why we use logarithmic smoothing in this work.

\section{Additional density profiles}
\label{app:density_profiles}
\begin{figure}
  \includegraphics[width=\columnwidth]{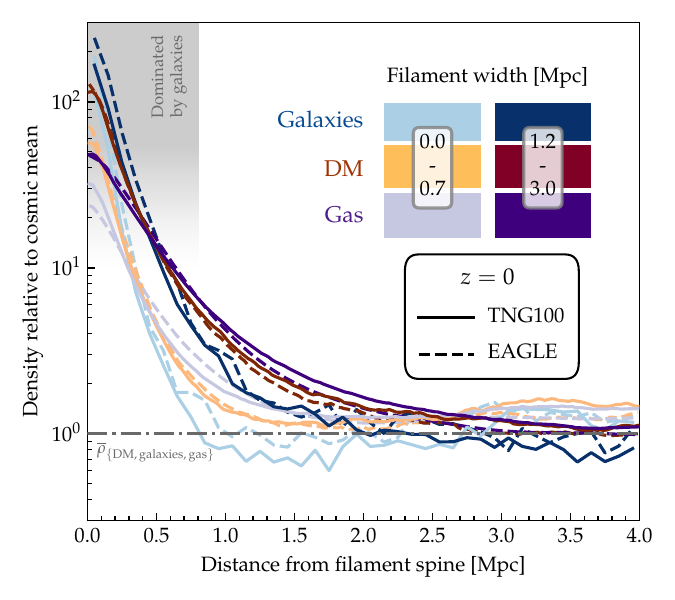}
  \caption{Density profiles around filaments in EAGLE and TNG100 at $z = 0$, in analogy to Fig.~\ref{fig:9_filament_density_profiles}. Galaxy, DM, and gas profiles are shown in blue, orange, and purple, each normalised to their respective cosmic mean. Light and dark shades represent the first and fourth quartile in filament width; for clarity we omit the two middle quartiles. Profiles for EAGLE-Ref100 and TNG100 are shown as dashed and solid lines, respectively. The grey shaded area marks overdensities $\gtrsim 10$ that are dominated by galaxies within the filaments. The two simulations agree qualitatively for all six profiles shown, and even quantitative discrepancies are minor.}
  \label{fig:B1_filament_density_profiles_eagle_tng_z0}
\end{figure}

\begin{figure}
  \includegraphics[width=\columnwidth]{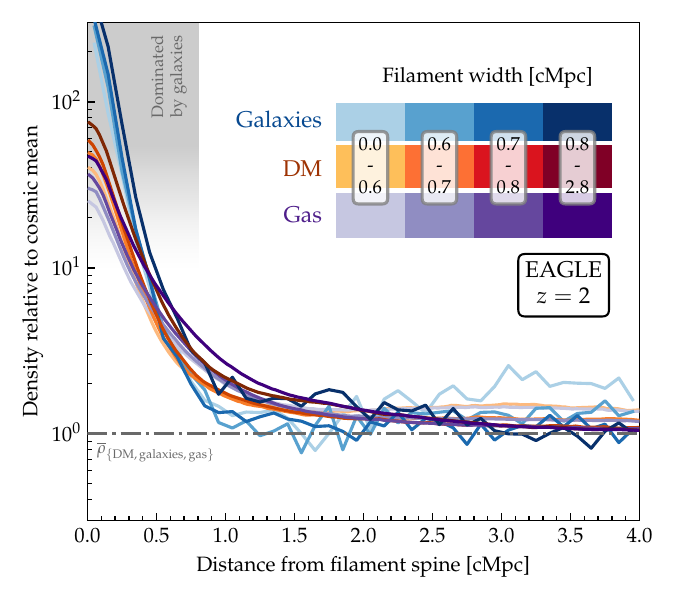}
  \caption{Density profiles around filaments at $z = 2$, in analogy to what is shown in Fig.~\ref{fig:9_filament_density_profiles} for $z = 0$. Galaxy, DM, and gas profiles are shown in blue, orange, and purple, each normalised to their respective cosmic mean (grey dash-dotted line). The grey shaded area marks overdensities $\gtrsim 10$ that are dominated by galaxies within the filaments. Different colour shades represent different quartiles in filament width. The profiles are qualitatively similar to those at $z = 0$, with some minor quantitative differences.}
  \label{fig:B2_filament_density_profiles_eagle_z2}
\end{figure}

For completeness, we include here the two additional sets of density profiles that were omitted from the main text as they do not show qualitatively new features. Fig.~\ref{fig:B1_filament_density_profiles_eagle_tng_z0} compares the $z = 0$ profiles around EAGLE and TNG100 filaments. The dashed lines for EAGLE are identical to those shown in Fig.~\ref{fig:9_filament_density_profiles}, but for clarity we only include the two extreme filament width quartiles. While there are some minor differences between the two simulations, these are small and do not affect any qualitative conclusions. In Fig.~\ref{fig:B2_filament_density_profiles_eagle_z2}, we show the density profiles for EAGLE at $z = 2$. Compared to the equivalent profiles for $z = 0$ as shown in Fig.~\ref{fig:9_filament_density_profiles}, they drop slightly faster (all profiles are close to the cosmic mean at 2 Mpc) but otherwise their qualitative features are the same as at $z = 0$. In both cases, overdensities $\gtrsim$10 correspond to galaxies within filaments so that the profiles in the central $\approx\,$500 kpc are not representative of the full filament length.

\end{appendix}

\end{document}